\documentclass[12pt,preprint]{aastex}

\newcommand{\symvec}[1]{\mbox{\boldmath${#1}$}}
\newcommand{\bvec}[1]{\textbf{#1}}

\shorttitle{A RINGLIKE DARK MATTER STRUCTURE IN Cl 0024+17}
\shortauthors{Jee et al.}

\begin{document}

\title{DISCOVERY OF A RINGLIKE DARK MATTER STRUCTURE IN THE CORE OF 
THE GALAXY CLUSTER Cl 0024+17} 

\author{M.J. JEE\altaffilmark{1}, 
        H.C. FORD\altaffilmark{1},
        G.D. ILLINGWORTH\altaffilmark{2}, 
	R.L. WHITE\altaffilmark{3}, 
	T.J. BROADHURST\altaffilmark{4},
        D.A. COE\altaffilmark{1, 5},  
        G.R. MEURER\altaffilmark{1}, 
        A. VAN DER WEL\altaffilmark{1},
	N. BEN\'{I}TEZ\altaffilmark{5},
	J.P. BLAKESLEE\altaffilmark{6}, 
	R.J. BOUWENS\altaffilmark{2}, 
	L.D. BRADLEY\altaffilmark{1}, 
        R. DEMARCO\altaffilmark{1}, 
	N.L. HOMEIER\altaffilmark{1},
        A.R. MARTEL\altaffilmark{1},
	AND 
        S. MEI\altaffilmark{1}}

\begin{abstract}
We present a comprehensive mass reconstruction of the rich galaxy cluster Cl 0024+17 at $z\simeq0.4$
from ACS data, unifying  both strong- and weak-lensing constraints.
The weak-lensing signal from a dense distribution of background galaxies ($\sim120 ~\mbox{arcmin}^{-2}$) 
across the cluster enables the derivation of a high-resolution parameter-free mass map. The
strongly-lensed objects tightly constrain the mass structure of the cluster inner region on an absolute scale,
breaking the mass-sheet degeneracy.
The mass reconstruction of Cl 0024+17 obtained in such a way is remarkable.
It reveals a ringlike
dark matter substructure at $r\sim75\arcsec$ surrounding
a soft, dense core at $r\lesssim50\arcsec$. 
We interpret this peculiar sub-structure as the result of a high-speed line-of-sight
collision of two massive clusters $\sim1-2$ Gyr ago. Such an event is also indicated by
the cluster velocity distribution. 
Our numerical simulation with purely collisionless particles demonstrates that
such density ripples can arise by radially expanding, decelerating particles that originally comprised
the pre-collision cores.   
Cl 0024+17 can be likened to the bullet cluster 1E0657-56, but viewed $along$ the collision axis at a much later epoch.
In addition, we show that the long-standing mass discrepancy for Cl 0024+17 between X-ray and lensing can be
resolved by treating the cluster X-ray emission as coming from a superposition of two X-ray systems.
The cluster's unusual X-ray surface brightness profile that requires a two isothermal sphere description supports this
hypothesis.

\end{abstract}

\altaffiltext{1}{Department of Physics and Astronomy, Johns Hopkins
University, 3400 North Charles Street, Baltimore, MD 21218.}
\altaffiltext{2}{UCO/Lick Observatory, University of California, Santa
Cruz, CA 95064.}
\altaffiltext{3}{STScI, 3700 San Martin Drive, Baltimore, MD 21218.}
\altaffiltext{4}{School of Physics and Astronomy, Tel Aviv University, Tel Aviv 69978, Israel.}
\altaffiltext{5}{Inst. Astrof\'{\i}sica de Andaluc\'{\i}a (CSIC), Camino Bajo de Hu\'{e}tor, 24, Granada 18008, Spain}
\altaffiltext{6}{Department of Physics and Astronomy, Washington State University, Pullman, WA 99164}

\keywords{gravitational lensing ---
dark matter ---
cosmology: observations ---
X-rays: galaxies: clusters ---
galaxies: clusters: individual (\objectname{Cl 0024+17}) ---
galaxies: high-redshift}

\section{INTRODUCTION \label{section_introduction}}
A galaxy cluster is still growing in today's Universe by continuously accreting other clusters/groups of galaxies.
Because the cluster and the infalling object are likely to reside in a common filament, their relative
motions are pre-dominantly expected to be one-dimensional (i.e., along the filament). This makes cluster cores the busiest places
in cluster dynamics subject to frequent near head-on collisions. Because the Universe is not old enough
to completely virialize these ever-growing clusters, many cluster cores are believed to
maintain bulk properties reflecting the cluster formation history even in the low-$z$ Universe. 

Recently, there have been a number of
reports on the detection of such clusters especially from X-rays (e.g., Mazzotta et al. 2001; Vikhlinin, Markevitch, \& Murray et al. 2001; 
Markevitch et al. 2002; Dupke \& White 2003; 
Henry et al. 2004; Belsole et al. 2005; Ferrari et al. 2006). Most of these clusters are characterized
by distinct discontinuities in temperature or density gradient of the intracluster medium (ICM).
However, these X-ray features hinting at the previous merging history are hard to identify if the collision is in progress
along the line-of-sight. Numerical simulations predict that in such a configuration the disruption of the ICM is azimuthally symmetric and thus 
the ICM of the cluster may appear spherically relaxed with radial temperature gradients (e.g., Takizawa 2000), which
cannot be easily disentangled from the intrinsic temperature gradient for a single relaxed system.
In the current paper we study one such case, namely Cl 0024+17, with gravitational lensing using deep Advanced
Camera for Surveys (ACS) observations.

Since its discovery by Humason \& Sandage (1957), the intermediate redshift ($z=0.4$) cluster Cl 0024+17 
has been a target of a number of studies (e.g., Zwicky 1959; Gunn \& Oke 1975; Koo 1988;
Kassiola et al. 1992; 
Bonnet, Mellier, \& Fort 1994; Smail et al. 1996;
Colley, Tyson, \& Turner 1996; Tyson, Kochanski, \& dell'Antonio 1998; Dressler et al. 1999; Broadhurst et al. 2000; Shapiro \& Iliev 2000; 
Soucail et al. 2000; Czoske et al. 2001; Kneib et al. 2003; Ota et al. 2004; Zhang et al. 2005; Metevier et al. 2006). One of the
most puzzling problems in Cl 0024+17 is the large mass discrepancy between the X-ray and the lensing results in the cluster core.
The five prominent arcs with a spectroscopic redshift of 1.675 (Broadhurst et al. 2000) at $r\sim30\arcsec$ from the cluster center
have prompted many authors to model the mass distribution in the central region (e.g., Colley et al. 1996; Tyson et al. 1998; 
Broadhurst et al. 2000; Comerford et al. 2006). Although the mass estimates and density profiles from these authors are at slight variance with
one another, it is evident that
the lensing analyses yield systematically higher core masses than the X-ray results by a factor of 3-4 (Ota et al. 2004; Zhang et al. 2005).
Because the five multiple images are well-resolved and the proposed mass models can successfully predict
the location, orientation, and parity of the five lensed images, the discrepancy has been attributed to problems
in X-ray mass estimation due to a severe departure from the hypothesized hydrostatic equilibrium.

One of the most convincing pieces of evidence for the recent dynamical disruption in the cluster center (i.e., 
responsible for the departure from the equilibrium)
comes from
the study of the cluster velocity field by Czoske et al. (2002). From $\sim300$ objects in the
redshift range $0.37<z<0.42$, they find that the redshift distribution is bimodal, showing two clear peaks
at $z=0.381$ and 0.395. By investigating the radial distribution of the velocity field of these two populations, 
they argue that the system has undergone a high-speed line-of-sight collision of two massive sub-clusters. 
Their numerical simulation with a mass ratio of about 2:1 reproduces the observed pattern of the velocity field.
If we are indeed observing a superposition of two clusters, the lensing mass should be the sum of the two components. On the other hand,
the X-ray mass estimation by the previous authors (e.g., Ota et al. 2004; Zhang et al. 2005) assumes
a single component although the collision scenario by Czoske et al. (2002) is acknowledged. The validity
of treating the X-ray emission as coming from a single X-ray systems, of course, 
depends on the state of the merger. If the two ICM systems have already merged and settled down to
an equilibrium state, the observed temperature and the slope of the profile can be viewed as representing 
the global properties of the relaxed post-merger system. However, if we are observing two post-collision clusters
that are still separated, the single component assumption should lead to a substantial underestimation of the projected mass.

Gravitational lensing has been claimed to be a unique method to measure mass properties of an object
free from any dynamical assumption. However, although it is true that the lensing signal does not
depend on the composition or the temperature of the deflector, in practice this ``assumption-free'' statement 
can be warranted
only when a sufficient number of lensing features are observable.

If the signal is sparse, a lensing mass reconstruction
inevitably necessitates some assumptions. Whether or not these assumptions are more dangerous than the quasi-equilibrium 
hypothesis in X-ray approaches 
certainly relies on the complexity of an individual system, as well as the number of observables. 
In this respect, mass reconstructions 
of Cl 0024+17 solely based on a limited number of multiple images are unwarranted despite the well-resolved morphology of the lensed images.
In principle, any curl-free vector field (i.e., gradient of a scalar potential) that correctly predicts the known multiple images can be 
suggested as a deflection field of
the system (surface mass density can be later derived by taking the divergence of this deflection field).
Because the location and shape of the source is unknown and the rest of the region not occupied by the multiple images
cannot be constrained, the solution is indeterminate. The situation can be slightly alleviated if some
physical considerations are used as additional constraints such as smoothness of mass distribution, resemblance
of dark matter distribution to cluster galaxy distribution, analytic behavior of density profiles (e.g., Navarro, Frenk \& White 1997), etc.
These assumptions are often implemented by placing parameterized dark matter halos on top of bright cluster ellipticals.
Nevertheless, the previous models in the literature (e.g., Tyson et al. 1998; Broadhurst et al. 2000; Comerford et al. 2006),
although all successful in predicting the multiple images, show somewhat discrepant mass distributions. 
This is not surprising because even with the help of those assumptions
one still faces many ambiguities such as where to place what types of dark matter halos.

Therefore, in the current investigation, we aim to present a parameter-free\footnote{
In this paper, we use the term ``parameter-free" or ``non-parametric'' mass reconstruction to refer to
a grid-based method. It is trivially obvious that the method also needs a set of ``parameters''
to define the grid.}
mass reconstruction of Cl 0024+17 combining 
multiply lensed images and the ellipticities of $\sim1300$ background galaxies. 
Several papers have already discussed this idea of unifying
strong- and weak lensing data in a parameter-free cluster mass reconstruction and 
applied the concept to observations  (e.g., 
Abdelsalem et al. 1998; Bridle et al. 1998; Seitz et al. 1998; Kneib et al. 2003; Smith et al. 2005; 
Bradac et al. 2005; Diego et al. 2005; Halkola et al 2006; Cacciato et al. 2006).
The approach used in the current study is similar to the ones investigated by 
Bridle et al. (1998), Seitz et al. (1998), and Bradac et al. (2005), 
who proposed to model a cluster mass distribution by setting up a two-dimensional grid over the cluster field.
We discuss the details of the method and the differences from the previous techniques
in \textsection\ref{section_theory}.

This approach utilizes information available in the entire cluster field
and is also assumption-free, completely blind to the distribution of the baryonic component of the cluster.
Because the ACS on board the {\it Hubble Space Telescope} ($HST$) can resolve very faint, but highly distorted distant
galaxies, the number density of the available source galaxies is unprecedentedly high. In addition, the deep,
six-passband coverage from F435W to F850LP provides secure photometric redshifts for individual
objects, allowing us not only to select the source population efficiently with a minimal dilution of the lensing signal
from non-background galaxies, but also to properly weight
their lensing efficiency according to their cosmological distances. 
The resulting mass map obtained in this
way will not be limited by a particular parameterization and will reveal any significant substructure
if the cluster core has indeed undergone a violent recent line-of-sight collision.

We organize our analyses as follows. In \textsection\ref{section_obs} we describe the observational aspects including basic data reduction, 
photometric redshift estimation, point-spread-function (PSF) correction, ellipticity measurement, etc. The basic theory and
algorithm of our parameter-free mass reconstruction are discussed in \textsection\ref{section_theory}. \textsection\ref{section_lensing_analysis} 
presents the result of the gravitational lensing analysis of Cl 0024+17. 
The result is discussed in \textsection\ref{section_discussion} before the conclusion in \textsection\ref{section_conclusion}. 

Throughout the paper we assume the $\Lambda$ CDM cosmology with $\Omega_M = 0.27$, $\Omega_{\Lambda}=0.73$, and
$H_0 = 71~\mbox{km s}^{-1} \mbox{Mpc}^{-1}$.
All the quoted uncertainties are at the 1 $\sigma$ level ($\sim68$\%).
We define the ellipticity as $(a-b)/(a+b)$, where $a$ and $b$ are the major- and minor-axes of the object, respectively.

\section{OBSERVATIONS \label{section_obs}}
\subsection{Data Reduction and Photometry \label{section_photometry}}
The intermediate redshift cluster Cl 0024+17 at $z=0.395$ was observed with the Wide Field Channel (WFC) of the
ACS in 2004 November as part of our Guaranteed Time Observations (GTO; ID 10325).
A single pointing ($\sim3\arcmin.3 \times \sim3\arcmin.3$ field of view) is centered 
at the cluster core ($\alpha_{2000}\simeq00^h:26^m:35^s, \delta_{2000}\simeq17\degr:09\arcmin:43\arcsec$) with
integrations of 6435, 5072, 5072, 8971, 10,144, and 16,328 s in the F435W, F475W, F555W, F625W,
F775W, and F850LP\footnote{These F435W, F475W, F555W, F625W,
F775W, and F850LP filters are commonly referred to as $B_{435}$, $g_{475}$, $V_{555}$, $r_{625}$, $i_{775}$, and $z_{850}$
, respectively. We follow this convention hereafter.}
filters, respectively. 
The low level CCD processing (e.g., overscan/bias subtraction, flat-fielding) 
was carried out using the STScI standard ACS calibration pipeline (CALACS; Hack et al. 2003) whereas
the final high-level processing (e.g., geometric distortion correction, cosmic-ray removal, image mosaicking)
was performed with the ACS GTO pipeline ``Apsis" (Blakeslee et al. 2003).
The integrity of the image alignment carried out by Apsis has been extensively tested in our
previous weak-lensing analyses (Jee et al. 2005a; 2005b; 2006; Lombardi et al. 2005). 
We used the Lanczos3 (windowed sinc function) kernel in drizzling (Fruchter and Hook 2002), which produces a sharper PSF and less
noise correlation between adjacent pixels than a ``square'' kernel (for
more detailed description of the noise and aliasing properties of Lanczos3 vs. other drizzle interpolation kernels see Mei et al. 2005).

In Figure~\ref{fig_cl0024}, we present the color-composite of the cluster field created from these final ``apsis'' products.
The image is displayed in the observed orientation (north is right and east is up) and is made square by trimming
the four sides of the original to match our mass reconstruction field.
The blue, green, and red intensities are represented with the $g_{475}$, $r_{625}$, and
$z_{850}$ fluxes, respectively. The well-known five multiple images at $z=1.675$ are labeled as A1-A5.
We also denote the two other multiple image system candidates 
(see \textsection\ref{section_implementation}) by B1-B2 and C1-C2.

Because ``drizzling" correlates pixel noise, one must use caution in producing the rms maps for
the final photometry. Apsis correctly calculates the rms for each pixel in the absence of this correlation.
We created a detection image by weight-averaging all bandpass images using their inverse variance maps.
Objects were detected with SExtractor by searching for at least five connected pixels above 1.5 times sky rms.
We manually removed $\sim330$ spurious objects (e.g., diffraction spikes around bright stars, uncleaned
cosmic-rays near the field boundaries). We note that some giant arcs and bright spirals
are undesirably fragmented and identified
as multiple objects by SExtractor. For these objects, we merged their segmentation map pixels by hand and
performed photometry by running SExtractor via the SExSeg software (Coe et al. 2006).

\subsection{Photometric Redshift and Selection of Background Galaxies}

Our deep, six-passband coverage of the cluster allows us to obtain secure photometric redshifts of objects 
in the Cl 0024+17 field. We used the isophototal magnitudes output by SExtractor to compute galaxy
colors and ran the revised Bayesian Photometric Redshift code 
(BPZ; Ben{\'{\i}}tez 2000; Ben{\'{\i}}tez et al. 2004) to determine their
photometric redshifts. The four spectral energy distribution (SED) templates of E, Sbc, Scd, and Im by Coleman, Wu, \& Weedman (1980), and the two
starburst templates of SB2 and SB3 by Kinney et al. (1996) are employed. 
As described in Benitez et al. (2004), the slopes of the SED were modifed to give agreement with the Hubble Deep Field (HDF)
spectroscopic redshifts. We also attempted to further calibrate the zero points of the photometry with
the publicly available spectroscopic redshift catalog of the Cl 0024+17 field (Moran et al. 2005). However,
we found that the estimated offsets are very small (we obtained 0.001, -0.012, 0.005, 0.013, 0.005 and -0.009
for the $B_{435}$, $g_{475}$, $V_{555}$, $r_{625}$, $i_{775}$, and $z_{850}$ filters, respectively) and the final result is not affected.
The post-launch sensitivity curves
of ACS by Sirianni et al. (2005) are used to obtain synthetic photometry of these templates
from $z=0.01$ to 6 at a redshift interval of $\Delta z =0.01$.
BPZ then compares this synthetic photometry to the observed photometry to determine the redshift probability
distribution for each galaxy. The strength of BPZ lies in its use of priors; each
galaxy is assigned a prior redshift probability distribution based on its magnitude.
This prior is multiplied to the probability obtained from the photometric SED fitting.
However, in these Cl 0024+17 images, the magnitudes of background objects are magnified by the cluster
lensing. Thus each background galaxy will be assigned a slightly biased prior.

To test the effects of the prior, we generated two sets of photometric redshift catalogs with and without priors obtained
from the Hubble Deep Field North (HDF-N) photometric redshift distribution. However, as demonstrated in Figure~\ref{fig_photo_z_compare}, 
we do not observe any systematic difference between these two sets except for the objects within the box.
The photometric redshifts of these objects obtained with HDF-N priors have a mean of $z\sim0.4$ with a dispersion of $\Delta z\sim0.2$ whereas
the photometric redshift estimation without priors identifies them as high-redshift objects at $2\lesssim z \lesssim 4$.
Many of these galaxies appear to have early-type morphology. This suggests that a substantial fraction of the population
is either associated with Cl 0024+17 or at lower redshifts, and their redshifts are correctly estimated only
with the help of priors.

The comparison of our photometric redshifts with the publicly available spectroscopic redshifts compiled by Moran et al. (2005) strongly
supports this hypothesis (Figure~\ref{fig_photo_z_spec}). The catalog contains 142 objects within our ACS field. Our photometric
redshifts with HDFN priors are consistent with the spectroscopic results as shown in the left panel of Figure~\ref{fig_photo_z_spec} whereas
the photometric redshifts computed without priors produce the catastrophic outliers at $2<z<4$ ({\it filled circle in the right panel}).

The reasons that these objects are mistaken for high-redshift objects are as follows. Because we estimate SEDs of galaxies
using broadband photometry without UV and near-infrared data, some degeneracies are inevitable. Especially, when there are some 
residual UV fluxes, the 4000 \AA~ break feature becomes weak and can be confused with other spectral features 
at high-redshifts; these degeneracies are worsened by photometric errors for faint galaxies. 
The typical shape of the redshift probability distribution of these objects has two dominant
peaks:one at $z\sim0.4$ and the other at $z=2\sim4$. In cases where the peak
at $z=2\sim4$ is greater than the peak at $z\sim0.4$, BPZ gives the former when
no prior is used. However, if the probability distribution is multiplied
by the prior and the $z\sim0.4$ peak now becomes greater of the two, the BPZ output is
$z\sim0.4$. Of course, the interpretation is that the object is ``too bright"
to be placed at $z=2\sim4$. 

Benitez et al. (2004) also showed in their photometric
redshift estimation with the WFPC2 $BVI$ photometry that without prior
the typical SED fitting method produces outliers at $z=2\sim4$.
The cloud of points at $z=2\sim4$ in the Figure 14 of Benitez et al. (2004) somewhat resembles 
the $z=2\sim4$ outliers in our paper (see also Figure 19 of Coe et al. 2006, which 
visually illustrates one such degeneracy in the SED fitting). 

Because no significant systematic discrepancy is found in other redshift ranges, we justify the
use of HDFN priors without any modification of the existing code. We show the photometric redshift
distribution of $\sim1820$ ($i_{775}<27.5$) non-stellar objects obtained with HDFN priors in the cluster field 
in Figure~\ref{fig_photo_z}.
The redshift spike at $z=0.4$ ({\it dotted line}) is clearly visible.

For our mass reconstruction of the cluster, we select objects whose photometric redshifts are greater than $z=0.8$ with
a minimum detection significance of $5\sigma$ at least in one filter.
These conservative values are chosen to ensure that the selection suffers from minimal contamination
by non-background population without reducing the number of usable galaxies substantially. The resulting
source catalog contains 1297 objects in the central $196\arcsec\times196\arcsec$ region of the cluster
($\sim120~\mbox{arcmin}^{-2}$).

\subsection{Ellipticity Measurements and Point Spread Function Corrections}

As in our previous weak-lensing analyses (Jee et al. 2005a; 2005b; 2006), we use adaptive elliptical 
Gaussian-weighted moments suggested by Bernstein \& Jarvis (2002)
to measure source ellipticities. The method has been extensively tested in Bernstein \& Jarvis (2002), 
Hirata \& Seljak (2003), Wittman et al. (2003), Park et al. (2004), 
Margoniner et al. (2005), Jarvis et al. (2003), etc.
This elliptical Gaussian weighting reduces the systematic underestimation of
the object ellipticities of the Kaiser, Squire, \& Broadhurst (1995, hereafter KSB) method, which uses a circular Gaussian
weighting. Previously, the elliptical Gaussian weighting was implemented by adaptively shearing 
object shapes to match the $circular$ Gaussian function in
shapelets (Refregier 2003; Bernstein \& Jarvis 2002). 

However, in the current paper, the implementation has been modified and
we now determine object shapes by directly fitting the PSF-convolved $elliptical$ Gaussian to the pixelized images. 
This method is conceptually identical to our previous shapelet approach, but provides some important
practical merits. The most significant advantage is that the method is better suited for
highly elongated objects, which cannot be
well-represented by shapelets. Shapelets are based on $circular$ Gaussian functions and thus introduce non-negligible aliasing 
for objects with high ellipticities. One such case is demonstrated in Figure~\ref{fig_aliasing}.
We note that, even for a moderately high order ($N=24$), the highly elongated shape of the object is not fully recovered 
and also
the decomposition creates some circular ripples around the object center. 
When we measure the ellipticity by directly fitting an elliptical Gaussian function to the object as proposed
in the current study, we
obtain $\epsilon=0.6311$, which is slightly higher than the $N=24$ shapelet measurement $\epsilon=0.6214$. 
The conventional KSB method based on a circular Gaussian function yields $\epsilon=0.4243$, substantially lower than
the shapelet or the elliptical Gaussian fitting methods.
The differences among these three measurements tend to increase
for higher ellipticity objects.

In principle, these aliasing features are
alleviated when the order of the shapelet decomposition becomes infinite. However, the convergence
is slow and unsatisfactory partly because the pixelization degrades the orthonormality of the basis functions. 
While this aliasing was not a problem in our previous weak-lensing analyses where not many objects have
such high ellipticities, it can create non-negligible biases in the current lensing study of Cl 0024+17, where
the ACS images provide numerous such arc(let)s within the field. 
In addition, for small objects whose effective radius approaches the size of the PSF, the PSF-convolved elliptical Gaussian fitting is
more numerically stable than our previous two-step solution with shapelets (i.e., ellipticity determination after deconvolution).

We create an object ellipticity catalog for each of the six filters and later combine the six catalogs to produce the final ellipticity catalog
by weighting each filter's output with its inverse variance; we do not observe any measurable 
systematic bias in ellipticity measurement
between different filters.
For each filter, a thumbnail image of an object is created and pixels belonging to other objects, if present, are masked out
using the SExtractor segmentation map.
The centroid of the object is determined from the detection image (see \textsection\ref{section_photometry}) and
is given as initial parameters. We freeze the background value using the SExtractor output, as this gives a more precise sky value than the
direct determination within the thumbnail image, especially when bright objects are nearby.
Then, elliptical Gaussian functions are convolved with the ACS PSF before being fitted to the object.
Our artificial shear test shows that the object ellipticities measured in this way reliably recover
the input shears (see Appendix A).

The PSF of ACS is time- and position-dependent as first noted by Krist (2003). We model the ACS PSF
using archival 47 Tuc observations as detailed in Jee et al. (2005a; 2005b; 2006). The position-dependent
PSF variation is conveniently described in $shapelets$ with coefficients varying as third order polynomials
of pixel coordinates.
Because the global
pattern changes depend on the $HST$ focus offset, we search the globular cluster images that were 
observed under a wide range of $HST$ focus values for the matching frames whose PSF patterns seem to fit those
in the Cl 0024+17 field. We use $\sim20$ stars in the Cl 0024+17 field in each filter as the pattern indicators, and
the fidelity of the PSF model is verified by checking the roundness of these stars after the Cl 0024+17 image is convolved with rounding
kernels created from the model. Figure~\ref{fig_psf}a and ~\ref{fig_psf}b show the PSF ellipticities
before and after the application of the rounding kernels, respectively.
The accuracy of our PSF model for the Cl 0024+17 field, judged from the reduction of the initial PSF anisotropy, is similar to the ones
in our previous studies (e.g., Jee et al. 2006), for instance giving a final mean ellipticity of $\epsilon=0.011\pm0.006$\footnote{Note that
in our previous papers, an ellipticity of a star was defined as $(a^2-b^2)/(a^2+b^2)$, which
is approximately a factor of two larger than the current definition $(a-b)/(a+b)$ for small values.} for F435W (and similar
values for other filters).

Occasionally, more than one template (star field) is good for the cluster field. Nevertheless, as long as the final mean 
residual ellipticity of the stars is $\sim0.01$ as in the case of the current paper, different templates do not make
a difference in our cluster mass reconstruction, where the lensing signal and the statistical errors from intrinsic ellipticity
overwhelm the PSF correction residual error. However, we suspect that even this $\sim0.01$ level
accuracy in PSF correction may be of concern to most cosmic shear studies in the future.

\section{A NON-PARAMETRIC MASS RECONSTRUCTION METHOD COMBINING STRONG AND WEAK-LENSING SIGNALS\label{section_theory}}
The concept of combining strong- and weak-lensing constraints to derive a cluster mass profile 
has been previously proposed and applied to observations (e.g., 
Abdelsalem et al. 1998; Bridle et al. 1998; Seitz et al. 1998; Kneib et al. 2003; Smith et al. 2004; 
Bradac et al. 2005; Halkola et al 2006; Cacciato et al. 2006). The approach used in the current study is similar to the method
of Seitz et al. (1998), who proposed to use individual galaxy ellipticities {\it without smoothing }
and to reconstruct the mass map through an entropy-regularized maximum-likelihood approach. The method
also suggests to construct the cluster's scalar potential first and to derive the mass map from this result.
This indirect derivation not only avoids the pitfalls of the finite-field inversion, but also enables an easy incorporation
of additional constraints such as strong-lensing features. Seitz et al. (1998) considered a case where
magnification information can be directly obtainable from the field. An important modification
of the method implemented here is to replace the magnification term in their likelihood function
with straightforward multiple image constraints. A similar modification was also proposed by Bradac et al. (2005).
However, they did not utilize the entropy of the mass to regularize the result, and their finite difference
scheme for the evaluation of $\kappa$ is different from ours.

Because there exist a number of excellent review papers on the subject (e.g., Kochanek 2004 for strong-lensing and
Bartelmann \& Schneider 2001 for weak-lensing), we summarize only
the basic lensing theory and equations in \textsection\ref{section_framework} necessary for the description
of our implementation (\textsection\ref{section_implementation}).

\subsection{Theoretical Frameworks \label{section_framework} }
Large cosmological distances between the observer, lens, and source galaxies justify
the ``thin'' lens approximation, where the mass distribution of the lens is only two-dimensional. Under this convenient
assumption, the deflection \symvec{\alpha} is handily expressed in terms of the following two-dimensional deflection potential:
\begin{equation}
\psi (\symvec{\theta})=\frac{1}{\pi} \int \mbox{d}^2 \theta^{\prime} 
\kappa (\symvec{\theta^{\prime}}) \ln |\symvec{\theta}-\symvec{\theta^{\prime}}| 
\label{eqn_potential}
\end{equation}
\begin{equation}
\symvec{\alpha}=  \nabla \psi. \label{eqn_alpha} \label{eqn_deflection}
\end{equation}
\noindent
In equation~(\ref{eqn_potential}), 
$\kappa$ is the surface mass density in units of the critical surface mass density $\Sigma_c=c^2 
D(z_s)/ (4\pi G D(z_l) D(z_l,z_s))$,
where $D(z_s)$, $D(z_l)$, and $D(z_l,z_s)$ are the angular diameter distance from the observer to the source,
from the observer to the lens, and from the lens to the source, respectively. The relation between $\kappa$ and $\psi$ can be
more compactly expressed using the gradient $\nabla$ operator as
\begin{equation}
\kappa=\frac{1}{2}\nabla^2 \psi \label{eqn_kappa}.
\end{equation}

The deflection \symvec{\alpha} 
(eqn.~\ref{eqn_alpha})
relates the source position \symvec{\beta} to the image position \symvec{\theta} via the lens equation:
\begin{equation}
\symvec{\beta}=\symvec{\theta} - \symvec{\alpha} (\symvec{\theta}). \label{eqn_lens}
\end{equation}
\noindent
The fact that \symvec{\alpha} is a function of the image position \symvec{\theta} implies that a single
position in the source plane can be imaged onto multiple locations. In a typical parametric strong-lensing modeling,
one uses equations~(\ref{eqn_potential})-(\ref{eqn_lens}) iteratively to construct the mass model $\kappa(\theta)$ that
correctly inverts all the sets of known multiple image positions to single source positions. In a weak-lensing regime
where the surface mass density is low (i.e., $\kappa < 1$), the system does not produce multiple images. However,
one can still detect coherent shape distortions of source galaxies. The Jacobian matrix describing the
distortion is obtained from the above lens equation (eqn.~\ref{eqn_lens}):
\begin{equation}
\textbf{A} \equiv \frac{\partial \symvec{\beta}}{ \partial \symvec{\theta}} = 
\left ( \begin{array} {c c} 1 - \psi_{11} & -\psi_{12} \\
                      -\psi_{12} & 1- \psi_{22}
          \end{array}   \right ) =
\left ( \begin{array} {c c} 1 - \kappa - \gamma _1 & -\gamma _2 \\
                      -\gamma _2 & 1- \kappa + \gamma _1
          \end{array}   \right ), \label{eqn_jacobian}
\end{equation}
\noindent	  
where the subscripts on $\psi_{i(j)}$ denote partial differentiation with respect to $\theta_{i(j)}$, and
the shears $\gamma_1$ and $\gamma_2$ are
\begin{equation}
\gamma_1=\frac{1}{2} (\psi_{11} -\psi_{22}) \label{eqn_shear1}
\end{equation}
\noindent
and
\begin{equation}
\gamma_2=\psi_{12}=\psi_{21} \label{eqn_shear2},
\end{equation}
\noindent
respectively. It is convenient to express the shears $\gamma_{1(2)}$
in complex notation: $\symvec{\gamma}=\gamma_1 + \bvec{i}\gamma_2$.
The Jacobian matrix (eqn.~\ref{eqn_jacobian}) transforms a circle into an ellipse with an
ellipticity $\hat{\bvec{g}}$:
\begin{eqnarray}
\hat{\bvec{g}}&=& \bvec{g}~~~~\mbox{if}~|\bvec{g}|<1 \nonumber \\
                          &=& \frac{1}{\bvec{g}^*}~~~~\mbox{if}~|\bvec{g}|>1,   \label{eqn_reduced_shear} \label{eqn_g_hat}
\end{eqnarray}			  
\noindent
where $\bvec{g}$ is the reduced shear $\bvec{g}=\symvec{\gamma}/(1-\kappa)$ (complex notation) and $\bvec{g}^*$ denotes the complex conjugate of 
$\bvec{g}$.
Note that the absolute value of $\hat{\mathbf{g}}$ above yields an ellipticity defined in the current paper.
 
Assuming that the intrinsic ellipticity distribution is isotropic, the mean ellipticity of galaxies
under the reduced shear $\bvec{g}$ is simply $\hat{\mathbf{g}}$ following the rule in Equation~(\ref{eqn_g_hat}). However,
many practical ellipticity measurements yield values systematically different from (in general lower than)
$\hat{\mathbf{g}}$. For example,
the conventional KSB method uses a circular Gaussian weighting in the measurement of the object
second moments and this choice of weighting circularizes object shapes, resulting
in the underestimation of the local shear (e.g., $\sim20$\% lower at $\hat{g} \sim 0.4$). 
Our shear estimation utilizing an elliptical Gaussian as a weighting scheme reduces such a 
systematic underestimation substantially as shown in Appendix A. Nevertheless, our numerical
simulation demonstrates that in a highly nonlinear
regime ($\hat{g}>0.4$) our ellipticity measurement still slightly underestimates the input shear partly
because the galaxies bend and become arclets. Potentially, this second order lensing effect (termed ``flexion'' 
in the Goldberg \& Natarajan (2002) paper)
can be utilized in the improvement of the local shear estimation (see also Massey et al. 2006 for
the suggestion of using shapelets for the flexion measurment).
In the current investigation, we
determine the correction factors by simulation and use them in our mass reconstruction.

So far, we have only considered a single source plane at a fixed redshift. In practice, source galaxies span
a wide range of redshifts, and the above equations must be scaled accordingly. We choose $z_f=3$ as 
the fiducial redshift and express the deflection potential (and the derived quantities) of the cluster 
with respect to the source plane at $z_f$. The translation of these physical values from $z_f$ to
a given redshift $z$ is straightforward. Using the ``cosmological weight" function 
\begin{equation}
W(z,z_f)=\frac{D(z_f) D(z_l,z)}{D(z_l,z_f) D(z)}, \label{eqn_cosmo_weight}
\end{equation}
\noindent
the surface mass density $\kappa$, shear $\gamma$, and deflection $\symvec{\alpha}$ are scaled as
\begin{equation}
\kappa(z)=W(z,z_f) \kappa(z_f)
\end{equation}
\begin{equation}
\symvec{\gamma}(z)=W(z,z_f) \symvec{\gamma}(z_f)
\end{equation}
\noindent
and
\begin{equation}
\symvec{\alpha}(z)=W(z,z_f) \symvec{\alpha}(z_f),
\end{equation}
respectively.
In addition, the reduced shear at a redshift of $z$ is now given as
\begin{equation}
\mathbf{g} (z) =  \frac {W(z,z_f) \symvec{\gamma}(z_f)} {1-W(z,z_f) \kappa(z_f)}   \label{eqn_mean_ellipticity}.
\end{equation}			     
\noindent
The expected mean ellipticity at a redshift of $z$ is then obtained by the rule in equation~(\ref{eqn_g_hat}).

\subsection{Implementation \label{section_implementation}}

We seek to construct a two-dimensional cluster deflection potential that correctly 
predicts the observed locations of the
multiple images and the ellipticity distribution of background galaxies in the ACS observations.
Because an observed lensing signal (i.e., shears and deflection) relates to a cluster mass only via a convolution, a direct
modeling of the $mass$ distribution within a finite field is subject to biases (masses
outside the field can affect the shears inside). Although one can attempt to alleviate this problem by extending
the field by a few factors, the scheme increases the number of unknown parameters substantially, causing
the minimization procedure to become prohibitively cumbersome. On the other hand, the deflection potential
can be locally converted to shears, deflection field, and mass density via 
equations~(\ref{eqn_deflection}),~(\ref{eqn_kappa}),~(\ref{eqn_shear1}), and~(\ref{eqn_shear2}). Therefore, we favor the direct estimation of the cluster lensing
potential as also advocated by many other authors 
(e.g., Bartelmann et al. 1996; Seitz et al. 1998; Bradac et al. 2005).
 
We set up a $52\times52$ potential grid over the central $210\arcsec\times210\arcsec$ region of Cl 0024+17.
At the inner $50\times50$ ($196\arcsec\times196\arcsec$) lattice points the shear $\gamma$, deflection field $\symvec{\alpha}$, and mass density $\kappa$ are
calculated by the central finite difference method. Note that we use the five nearest points as in Seitz et al. (1998) to evaluate 
$\kappa$ whereas
Bradac et al. (2005) used the four additional diagonal points (a total of nine).
 
Then, these values at lattice points are bicubic interpolated 
to estimate the lensing observables at each galaxy location. We find that the bicubic interpolation 
provides not only a smoother result, but also smaller $\chi^2$ values than the bilinear interpolation 
although
the evaluation is computationally much more expensive. Particularly, we notice that 
the bicubic interpolation substantially
outperforms the bilinear interpolation in the deprojection of the multiply lensed objects.

When evaluating $\gamma_2$ at the four corners of the $50\times50$ grid, 
we use the finite difference scheme of eqn. 25.3.27 in Abramowitz \& Stegun (1984), following the suggestion 
of Seitz et al. (1998). This prevents the four corners of the $52\times52$ potential grid from
being used in the minimization below. 

We now desire to find a set of parameters describing the cluster potential by minimizing the following function:
\begin{equation}
f = \frac{1}{2}\chi^2_{\mu} + \frac{1}{2}L_{\epsilon} + R, \label{eqn_f} \label{eqn_minimization}
\end{equation}
\noindent
where $\chi^2_{\mu}$ is the dispersion of multiple images in the source plane, $L_{\epsilon}$
is the log-likelihood function for the shear, and $R$ is the regularization
term that is required to prevent the minimization procedure from overfitting the data.
The factor $1/2$ in the first and second term is included to
ensure that the posterior probability distribution is proportional to $\mbox{exp}[-(f-f_m)]$ (i.e., without any
additional factor in front of $f$), where
$f_m$ is the value of equation (\ref{eqn_minimization}) at the global minimum.

We define $\chi^2_{\mu}$ for a single source at a redshift $z$ with $M$ multiple images and
$N$ $knots$ as
\begin{equation}
\chi^2_{\mu} = \sum_{m=1}^{M} \sum_{n=1}^{N} \frac{\left ( \symvec{\theta}_{m,n}  - W(z,z_f) \symvec{\alpha}(\symvec{\theta}_{m,n}) -
 \symvec{\beta}_n\right )^2}{\sigma_{m,n}^2} \label{eqn_chi_m}
\end{equation}
\begin{equation}
\symvec{\beta}_n =  \frac{1}{M}\sum_{m=1}^{M} \left ( \symvec{\theta}_{m,n}  - W(z,z_f) \symvec{\alpha}(\symvec{\theta}_{m,n}) \right ) 
\label{eqn_multi}
\end{equation}
\noindent
where $\symvec{\theta}_{m,n}$ and $W(z,z_f)\symvec{\alpha}(\symvec{\theta}_{m,n})$ are 
the coordinate of 
the $n^{th}$ knot of the $m^{th}$ multiple image and the scaled deflection at its redshift $z$, respectively.

The choice of $\sigma_{m,n}$ is important. Using a fixed value throughout the minimization biases
the model toward high magnification because the numerator of equation~(\ref{eqn_chi_m}) decreases as the magnification increases
regardless of the goodness of the agreement of individual knots. This often leads to incorrectly small $\chi_{\mu}^2$
values (thus also unreasonably small error estimates for the fitting parameters).
One possible solution is to utilize the magnification
tensor $M$ to scale the error according to the magnification. However, this scheme becomes numerically very unstable and $\chi^2_{\mu}$
diverges if any of the multiple images are close to critical curves. In the current implementation, we developed a novel, simple scheme, 
which does not
bias the model toward high magnification but without the use of the magnification tensor. We normalized the coordinates
of the knots of the multiple images in the source plane in such a way that they always range between zero and one. 
These normalized coordinates change as we iterate. Then, 
a fixed value of the uncertainty becomes a fractional uncertainty and no longer biases the model
toward high magnification. Nevertheless, there is one concern about this normalization.
A solution where the source positions have a couple of outliers (i.e., setting the $0-1$  scaling)
plus a cluster of points with small scatter can also yield low $\chi^2_\mu$ values. However,
this configuration is highly disfavored in practice by the following two reasons. First, the weak-lensing signals tend to keep the solution
away from the mass model that predicts such an unusual source configuration. Second, once all the source positions are
``locked'' closely, further iterations do not produce such outliers because a source position ``drifting'' away
from the rest of the source locations increases the above $\chi_\mu$ steeply. 
The exact value of $\sigma_{m,n}$ is not critical, but should not be set too small as the model is limited by
the finite resolution of the grid and interpolation errors. 

It is appropriate to point out here that the above source plane minimization can be potentially replaced with
an image plane minimization (Kochanek 2004), where the predicted source positions are compared with
their observed positions in the image plane. This scheme would obviate the need to rescale the $\sigma_{m,n}$
values as required in the current source plane minimization. Furthermore, the minimization would
disfavor the model that predicts any unobserved image because it can now penalize the resulting $\chi^2_\mu$ values.
Unfortunately, the image plane minimization is computationally much more expensive, requiring a nontrivial 
image plane search. Because our system already involves the time-consuming numerical estimation of the function 
derivatives (see below), this image plane minimization is highly unfeasible if not impossible in the
current approach. However, we emphasize that the above source plane minimization with renormalization of
the source plane coordinates prevents the model from being biased toward high magnification. The only remaining
concern is occasional prediction of unobserved multiple images. Nevertheless, the presence of these unobserved
multiple images should not discredit the model because, as discussed in \textsection\ref{source_image_reconstruction},
the converged potential can be always modified manually to remove those spurious objects by making some negligible
changes to the model. 

We used the well-known five multiple images from the single source at $z=1.675$ and
two additional multiple system candidates in the evaluation of equation~(\ref{eqn_chi_m}).
The location of these two new systems in the cluster field are denoted as B1-B2 and C1-C2 in Figure~\ref{fig_cl0024} and
their cutout images are shown in Figure~\ref{fig_new_multi}. In theory, the total number of multiple images in an extended lens is always odd as long as the inner mass profile
is shallower than the singular isothermal (i.e., $\propto r^{-2}$) profile. Nevertheless, it is
well known that many observed lens images seem to have two or four images because
the third or fifth image is usually either much fainter or obscured by bright galaxies.
In the current cluster Cl 0024+17 we also have not yet found any convincing third image candidate for 
the B1-B2 and C1-C2 systems. The B1-B2 system is originally identified
by Broadhurst et al. (2000) with WFPC2 observations. The photometric redshift
of the system is $\bar{z}_{phot}\simeq 1.3$. Our initial mass model based on this B1-B2 system along with the A1-A5 system
predicts that the C1-C2 images with $\bar{z}_{phot}\simeq 2.8$ are also multiply lensed. We impose relatively
loose constraints to the convergence of these two systems, considering the typical, large uncertainty ($\delta z\sim0.1$) of the photometric
redshift estimation and the lack of morphological features of these systems.
We set $\sigma_{m,n}$ to 0.03 for A1-A5 and to 0.3 for B1-B2 and C1-C2; an order
of magnitude larger $\sigma_{m,n}$ values are used for B1-B2 and C1-C2. These values are empirically determined 
in our attempt to make the above $\chi_{\mu}^2$ per degree of freedom become close to unity.

The individual galaxies in A1-A5 multiple system are well resolved, and we choose 10 bright knots for each source as inputs
to Equation~\ref{eqn_multi} ($10\times5\times2=100$ constraints). Although Tyson et al. (1998) claim that 
they can characterize each source with 58 smooth disks (four parameters for each disk) in their modeling, 
our experiments demonstrate that increasing the number of constraining 
features further does not improve the model; because the deflection field varies very smoothly, excessively additional 
constraints provide only redundant information. The B1-B2 and C1-C2 systems possess relatively unclear morphology
and thus we characterize each source with only four positions ($4\times4\times2=32$ constraints).

The log-likelihood function $L_{\epsilon}$ can be derived from the assumption that the ellipticity distribution
for the presence of the reduced shear $g$ is a Gaussian:
\begin{equation}
p_{\epsilon} (\epsilon|g) \propto \frac{1}{\pi \sigma_{\epsilon}^2(\hat{g})} 
\mbox{e}^{ -  |\symvec{\epsilon}-\hat{\mathbf{g}}|^2/ \sigma_{\epsilon}^2(\hat{g})}. \label{eqn_ellipticity_distribution}
\end{equation}
\noindent
Although we know that the exact shape of the lensed ellipticity distribution is not Gaussian, 
equation~(\ref{eqn_ellipticity_distribution}) is a convenient approximation,
which represents the first and second moments of the lensed ellipticities (Geiger \& Schneider 1999).

Then, the log-likelihood function for $K$ background galaxies is given as
\begin{equation}
L_{\epsilon}= \sum_{k=1}^K \left ( \frac{ |\symvec{\epsilon_k} - \hat{\mathbf{g}}|^2 } {\sigma_k^2(\hat{g})} +
 \ln{\sigma_k^2(\hat{g})} \right ), \label{eqn_log-likelihood}
\end{equation}
\noindent
where $\sigma_k (\hat{g})$ is the ellipticity dispersion for the $k^{th}$ galaxy under the influence
of the shear $g$ and can be approximated by adding the intrinsic ellipticity dispersion and the measurement error in
quadrature:
$\sigma_k^2(\hat{g})=\sigma_{\epsilon}^2(\hat{g}) + \sigma_{k,err}^2$.
The ellipticity dispersion for a given $\hat{g}$ is often assumed to follow
the simple analytic form:
\begin{equation}
\sigma_{\epsilon}(\hat{g})=\sigma_{\epsilon} (0) (1-\hat{g}^2) , \label{eqn_sigma_dispersion}
\end{equation}
\noindent
where $\sigma_{\epsilon} (0)$ is
the intrinsic ellipticity dispersion of the source population in the absence of gravitational lensing.
From our artificial shear test, we find that Equation~(\ref{eqn_sigma_dispersion}) with $\sigma_{\epsilon} (0)\sim0.3$ 
is a good approximation over a wide range of $\hat{g}$. Nevertheless, it slightly underestimates the true
dispersion at low $\hat{g}$ and overestimates the value at high $\hat{g}$. 
Therefore, we attempt to obtain a better analytic expression and find that the relation
\begin{equation}
\sigma_{\epsilon} (\hat{g})= 0.31~(1-\hat{g}^2)^{1.11} \label{eqn_sigma_dispersion_better}
\end{equation}
\noindent
provides a better fit to the simulation result (see Appendix B).

Finally, we need to define the regularization $R$, which governs the overall smoothness of the mass reconstruction. The need for
this regularization $R$ is obvious when we compare the number of free parameters 2697 (see below) with the number of the constraints
2726 ($1297 \times 2$=2594 from the ellipticities and 132 from the multiple images). Without the regularization, the minimization
will overfit the data unless the number of constraints is significantly larger than the number of parameters.
We adopt the 
following maximum-entropy
regularization (Press et al. 1992) implemented by Seitz et al. (1998):
\begin{equation}
R=\eta \sum_{p,q} \hat{\kappa}_{p,q} \ln \left ( \frac{\hat{\kappa}_{p,q}}{b_{p,q}} \right ), \label{eqn_regularization}
\end{equation}
\noindent
where $\hat{\kappa}_{p,q}$ and $b_{p,q}$ are the surface mass density and prior at the grid point $(p,q)$, respectively, normalized
in such a way that the summation over the entire grid becomes unity. The maximum-entropy method (MEM) leads
to a mass reconstruction as smooth as possible while preserving details that the data constrain.
The parameter $\eta$ is to control the smoothness
of the resulting mass map and needs to be adjusted in such a way that
$\chi^2$ per degree of freedom remains close to unity. 

As suggested by Seitz et al. (1998), one can use the result of the direct mass
reconstruction as an initial prior and update it at the beginning of the next minimization step
with the smoothed version of the previous mass map. However, the choice of the initial prior does not
determine the final result and
one can start the minimization with a flat prior to reach the virtually identical final result, but of course with
many more iterations. 
As mentioned above the four corner points of the $52\times 52$ grid do not enter the minimization. In addition, because the
zero point of the deflection potential and the translation of the source plane are arbitrary, we need
to fix three additional grid points. Therefore, the number of free parameters is $(50+2)\times (50+2)-7=2697$.

We choose the Davidon-Fletcher-Powell (DFP) algorithm (Press et al. 1992) as
our main optimization scheme to minimize the target function (eqn.~\ref{eqn_minimization}).
The DFP algorithm constructs an inverse Hessian matrix iteratively and uses it along with the partial derivatives
of the function to determine the next iteration point $\symvec{\psi}$:
\begin{equation}
\symvec{\psi} - \symvec{\psi}_i = - \bvec{A}^{-1} \cdot \nabla f(\symvec{\psi}_i).
\end{equation}
\noindent
The complexity of our target function (i.e., the use of bicubic interpolation,
maximum-entropy regularization, etc.) makes it non-trivial, if not impossible, 
to write the derivatives $\partial f / \partial \psi_k$
in linear terms of $\psi_k$. Therefore, we feed numerically calculated derivatives to the IDL
implementation of the algorithm (DFPMIN). Although the above algorithm is efficient, we
find that the minimization occasionally gets stuck in local minima. Hence, we complement the minimization
procedure with the gradient-free Direction-Set method (Press et al. 1992). In general,
this minimization scheme, which does not require explicit evaluation of the gradients
of the target function, converges slower than the DFP method above or other gradient-based
techniques (e.g., conjugate-gradient method). However, we observe that this Direction-Set method
more effectively resolves the local minimum. Consequently, we restart the minimization
with this second algorithm whenever the DFP minimization converges to local minima.
Although it is in general extremely difficult to reach a unique set of parameters for
a large system in a strictly mathematical sense, we are convinced that our final set
of parameters are very close to the $true$ global minimum. We examine the quasi-uniqueness
of our solution in two ways. First, we repeat the minimization with different choices of initial
conditions (and priors) and verify that they all lead to $virtually$ identical results.
Second, we perturb the converged set of parameters by adding small random numbers and execute the minimization
with this new set of parameters. No significant drifts from the original set of parameters are observed.

\subsection{Uncertainties of the Mass Reconstuction \label{section_uncertainty}}

The proper interpretation of the features in the mass reconstruction necessitates our
understanding of the noise properties. In general, the complex relation between lensing
observables and the derived mass map makes the noise estimation non-trivial. 
In the current study, we choose to estimate the uncertainties utilizing the Hessian
matrix of the target function $f$ (eqn.~\ref{eqn_minimization}) at the location of the minimum $\hat{\symvec{\psi}}$.
When $\symvec{\psi}$ is sufficiently close to the location of the function minimum $\hat{\symvec{\psi}}$, the target function $f$
can be approximated by a quadratic form:
\begin{equation}
f(\symvec{\psi})\simeq f(\hat{\symvec{\psi}}) + (\symvec{\psi} - \hat{\symvec{\psi}} ) \cdot \nabla f(\hat{\symvec{\psi}}) +
\frac{1}{2} ( \symvec{\psi} - \hat{\symvec{\psi}} ) \cdot \bvec{A} \cdot ( \symvec{\psi} - \hat{\symvec{\psi}} ), \label{eqn_quadratic}
\end{equation}
\noindent
where the first order term vanishes because $\nabla f =0$ at $\hat{\symvec{\psi}}$.
By exponentiating the above equation, we get the posterior distribution:
\begin{equation}
P(\symvec{\psi}) \propto \exp \left[ - \left ( f(\symvec{\psi})-f(\hat{\symvec{\psi}}) \right )  \right ] \propto
\exp \left [-\frac{1}{2}  ( \symvec{\psi} - \hat{\symvec{\psi}} ) \cdot \bvec{A} \cdot ( \symvec{\psi} - \hat{\symvec{\psi}} ) \right ].
\end{equation}
\noindent
That is, the posterior distribution $P(\symvec{\psi})$ becomes Gaussian in the neighborhood of $\hat{\symvec{\psi}}$
with a covariance matrix being the inverse of the Hessian $\bvec{A}^{-1}$.
Bridle et al. (1998) show that this Gaussian approximation agrees with the result from their Monte Carlo experiments.
Of course, because we estimate the deflection potential $\psi$ (not the convergence $\kappa$ directly) in our study, it is necessary
to propagate the errors, accordingly.

\section{GRAVITATIONAL LENSING ANALYSES \label{section_lensing_analysis}}

\subsection{Mass Reconstruction and Discovery of a $r\simeq0.4$ Mpc ringlike Dark Matter Structure \label{section_massmap}}
Our mass reconstruction of Cl 0024+17 is presented in Figure~\ref{fig_massmap}.
Figure~\ref{fig_massmap}a shows the $50\times50$ mass map derived from the converged $52\times52$ deflection potential.
The corresponding rms map (Figure~\ref{fig_massmap}b) is derived from the Gaussian approximation (\textsection\ref{section_uncertainty}).
We note that the rms map yields, a mean uncertainty of $\bar{\kappa}\sim 0.02$, giving
higher values at the field boundary and lower values for the region constrained by the strong-lensing data.
Also displayed is the bicubic interpolated version (Figure~\ref{fig_massmap}c) with a streched color table
to emphasize the low-contrast feature. This map was reproduced with a slightly larger regularization constraint
(the final ellipticity $\chi^2$ per galaxy is $\sim1.3$).

The mass map reveals the striking sub-structure of Cl 0024+17,
characterized by the soft, high density core at $r\lesssim50\arcsec$ 
and the moderately overdense, ringlike substructure 
at $r\sim75\arcsec$ (see also Figure~\ref{fig_massmap}d).
The ringlike substructure is strongly
constrained by the weak-lensing signals at $r\gtrsim50\arcsec$ and appears even when the mass reconstruction is performed
$without$ the strong-lensing data. The feature can be also clearly identified
in the radial density profile (\textsection\ref{section_radial_mass}) and in the tangential shear profile
(\textsection\ref{section_tangential_shear}). 
In the absence of the ``ring,'' the mean mass density in the annulus ($r=65\arcsec-85\arcsec$) is $\kappa\sim0.65$.
With respect to this ``background'' density, the feature is significant at $\gtrsim 5~\sigma$ levels.
Because of the finite number and the nonuniform distribution of background galaxies,
the ringlike feature
is lumpy on a small scale. We also note that there is somewhat large-scale
azimuthal variation on the structure, which is discussed in the context of the origin
of the feature in \textsection\ref{section_nbody}.

The mass peak is in good spatial agreement with the 
giant elliptical galaxies in the cluster center and is elongated in the direction defined by the
three co-linear galaxies (see Figure~\ref{fig_xraymap}a and~\ref{fig_xraymap}b). Moreover, this mass peak 
coincides with the X-ray peak first revealed by $Chandra$, which also appears to possess an elongation
in the same direction (Figure~\ref{fig_xraymap}c and~\ref{fig_xraymap}d). By a careful comparison however, we note that
the X-ray centroid is offset to the north-east by $\sim 10\arcsec$ and is close to the galaxy \#380 
($\alpha_{2000}\simeq00^h:26^m:36.03^s, \delta_{2000}\simeq17\degr:09\arcmin:45.9\arcsec$) whereas
the mass centroid is near the galaxy \#374 ($\alpha_{2000}\simeq00^h:26^m:35.69^s, \delta_{2000}\simeq17\degr:09\arcmin:43.12\arcsec$);
in referring to the galaxies we use the object IDs defined in the catalog
of Czoske et al. (2002).  

Broadhurst et al. (2000) modeled the cluster strong-lensing mass by placing eight $circular$ Navarro-Frenk-White (NFW) halos on top
of bright elliptical galaxies. Their mass map also shows that the three co-linear galaxies mentioned above
define the mass peak. Comerford et al. (2006) were also able to reproduce the five multiple images yet only including those
three elliptical galaxies, modeling them as three $elliptical$ NFW halos. Because the presence of the fifth
image (i.e., denoted A5 in Figure~\ref{fig_cl0024}) strongly
constrains the location of the mass peak, it is not unexpected to observe the good agreement in the location of the mass peak
among the different lensing studies. 

\subsection{Radial Mass Profile \label{section_radial_mass}}

From the two-dimensional mass map in Figure~\ref{fig_massmap}, we can infer that the mass distribution
is nearly axisymmetric and also the projected density does not decrease in a monotonic manner as a function of radius. 
The soft core is surrounded by a low-density annulus at $r\sim50\arcsec$ and then by moderately high-density
ringlike structure at $r\sim75\arcsec$.
Here we examine
the radial mass density profile of the Cl 0024+17 in detail. We choose the geometric center of 
the ringlike structure ($\alpha_{2000}\simeq00^h:26^m:35^s.92,~\delta_{2000}\simeq17\degr:09\arcmin:35\arcsec.5$) as the cluster center 
to calculate the azimuthally averaged density profile 
(Figure~\ref{fig_radial_density}). 
The core and $ring$ revealed in our previous two-dimensional mass reconstruction are also visible in
this radial mass density plot.
The projected density of the cluster flattens outside the core 
and creates the $bump$ at $r\sim75\arcsec$. The shape of the mass profile in this region
is strongly constrained by the weak-lensing signals; the feature appears as a $trough$ in the
tangential shear profile (see \textsection\ref{section_tangential_shear}).
The dotted lines represent the 1 $\sigma$ deviation of the azimuthal mean, which reflect
not only the noise level, but also the intrinsic azimuthal variation. Because of the off-centered
mass peak, the azimuthal deviation is large at small radii ($r\lesssim 20\arcsec$).
In \textsection\ref{section_massmap} we estimated the significance of the ringlike structure to be at least 5 $\sigma$ 
from the two-dimensional mass profile and the derived rms map. 
In this one-dimensional profile, the significance of the bump in the $r=60\arcsec-85\arcsec$
region with respect to the azimuthal mean ($\bar{\kappa}\simeq0.65$) at the trough ($r\sim55\arcsec$) or 
the ``tail'' ($r\sim90\arcsec$) is $\sim 8\sigma$.
 
The overall shape of this radial density profile looks more striking when compared
to the results of the previous studies (Figure~\ref{fig_density_compare}).
We transformed the results of Tyson et al. (1998) (dotted line), Broadhurst et al. (2000) (dashed line),
and Ota et al. (2004) (dot dashed) using the current cosmological parameters.
A significant difference among the models is undeniable. As already indicated by Ota et al. (2004),
the X-ray mass is far less than the other three lensing results; a more recent X-ray
analysis with XMM-Newton (Zhang et al. 2005) (omitted here) yields even slightly lower values.
The low core densities ($\kappa < 1$) predicted by these X-ray analyses violate
the fundamental condition of the strong-lensing, which requires a projected mass
density greater than unity in the cluster core.

The result of Tyson et al. (1998) gives the highest core density at $r\lesssim 15\arcsec$, but
the lowest density at large radii $r \gtrsim 20 \arcsec$. Because the location of the
critical curves at a fixed redshift are invariant under the transformation $\kappa \rightarrow \lambda \kappa + (1-\lambda)$,
a strong-lensing modeling based on single redshift multiple images is subject to this mass-sheet degeneracy.
Therefore, the two lensing mass profiles of Tyson et al. (1998) and Broadhurst et al. (2000) inside the Einstein radius
will
roughly overlap each other under the above transformation with a proper choice of $\lambda$.
However, this degeneracy is lifted in our result with the help of the two added source planes at $z=1.3$ and $2.8$.
Moreover, the weak-lensing data extended to the critical regime provide
additional constraints in resolving the degeneracy because the arc(let)s whose shears approach $g\sim1$ inform us of
the redshift-dependent, critical curve locations.

The mass profile of Broadhurst et al. (2000) 
is, nevertheless, somewhat similar to our mass profile at $r \lesssim 30\arcsec$, but the difference is observed
at larger radii. Our mass profile outside the Einstein radius is tightly constrained by the weak-lensing data.
A detailed comparison between the model prediction and the observed shear profile
is in \textsection\ref{section_tangential_shear}.

The flat density profile of our model implies that projected cumulative masses rise steeply 
(Figure~\ref{fig_projected_mass}). The projected mass within 
the radius of the $z=1.675$ arc ($r\simeq 30\arcsec$) is often quoted for mass comparison
between different mass models.
Our model predicts a projected mass of $M(r<30\arcsec)=(1.79\pm0.13)\times10^{14} M_{\sun}$,
which is consistent with the result of Broadhurst et al. (2000); when we reproduce their
model in the current cosmology, we obtain $M(r<30\arcsec)\simeq1.84\times10^{14} M_{\sun}$.
The result in Tyson et al (1998) was obtained assuming the $\Omega_M=1$ flat
Universe. Nevertheless, when only the difference in $h$ is considered in the transformation of the result, 
it also gives a similar mass
of $M(<30\arcsec)\sim1.6\times10^{14} M_{\sun}$. This excellent agreement of the projected total
masses within the radius of the arcs among different models is not surprising, however, because for an axisymmetric
lens the mean mass density within an Einstein radius becomes unity regardless of the difference in the radial profile.

Ota et al. (2004) claim that their X-ray mass from the $Chandra$ analysis is smaller than the lensing result
by a factor of 3 at $r=35\arcsec$ in the $\Omega_M =1$ flat Universe.
The difference becomes somewhat reduced if the result is reproduced in the current cosmology and
a slightly smaller aperture $r=30\arcsec$ is chosen.
Under the hypothesis of hydrostatic equilibrium, the $Chandra$ X-ray measurements
of $\beta=0.71$ and $T=4.47$keV from Ota et al. (2004) imply $M(r<30\arcsec)\sim7.94\times 10^{13} M_{\sun}$,
which is still lower than the lensing estimation by a factor of 2.

\subsection{Reduced Tangential Shears \label{section_tangential_shear}}

The reduced tangential shear is defined as
\begin{equation}
 g_T  =   < - e_1 \cos 2\phi - e_2 \sin 2\phi > \label{tan_shear},
\end{equation}
\noindent
where $\phi$ is the position angle of the object with respect to the cluster center, and
$e_{1(2)}$ is the object ellipticity. Because each galaxy has its own 
intrinsic shape, the reduced shear can be estimated by taking azimuthal averages
in radial bins $\Delta r$.
Figure~\ref{fig_tan_shear} shows the reduced tangential shears of Cl 0024+17 after the systematic
underestimation is corrected for high shears (see Appendix A).
The overall shape of the profile is consistent with our expectation for a typical axisymmetric lens.
The background galaxies are most strongly stretched in the tangential direction near the Einstein radius
and the tangential shears decrease for increasing $r$. Inside the Einstein radius, the tangential shear similarly
goes down as $r$ decreases. However, near the cluster center, the lensed images tend to become radially stretched. Hence
the observed shears must cross the zero line and become negative. Filled squares are the results from the 45 \degr rotation ($B$-mode)
test. The lensing signals disappear and the residual amplitudes are consistent with the statistical errors.

Also shown in Figure~\ref{fig_tan_shear} are the predicted tangential shears from the Broadhurst et al. (2000) model (dashed line) and
our result (solid line). These predicted values are estimated by placing circular objects at the location of the background
galaxies; assuming the intrinsic ellipticity dispersion (eqn.~\ref{eqn_sigma_dispersion_better})
the expected errors of these points are similar to those of the observed points, but are omitted for readability.
The Broadhurst et al. (2000) model predicts tangential shears
consistent with our observation at $r\lesssim40\arcsec$, but much higher at $r\gtrsim40\arcsec$.
The discrepancy implies that their mass profile inside the Einstein radius is similar to our result, but
steeper at larger radii; we already noticed this 
directly in the comparison of the radial density plot (Figure~\ref{fig_density_compare}).
As the strong-lensing region used by Broadhurst et al. (2000) is limited only to the
region interior to the Einstein radius ($r \lesssim 30\arcsec$),
their predicted shear profile is increasingly discrepant for $r>40\arcsec$.

An interesting feature in this tangential shear plot is the dip present at $r\sim75\arcsec$. It is
clear that this feature
reflects the ringlike sub-structure seen in the two-dimensional mass map (Figure~\ref{fig_massmap})
or the $bump$ in the radial density plot (Figure~\ref{fig_radial_density}).
We stress that, as these data points are uncorrelated, the observed departure from a monotonic decrease is
highly significant.
It may not be intuitive, however, to understand why the $bump$ in the radial density profile appears as a $dip$ in
the tangential shear profile as the relation between mass density and reduced shear is complicated.
For an axisymmetric lens,
the reduced shear at $r$ is given by 
\begin{equation}
g(r)=\frac{\bar{\kappa}(<r)-\kappa(r)}{1-\kappa(r)}, \label{eqn_shear_mass}
\end{equation}
\noindent
where $\bar{\kappa}(<r)$ is the average surface density within $r$. Using Equation~(\ref{eqn_shear_mass}) along
with the radial density plot (Figure~\ref{fig_radial_density}), it is possible to
qualitatively reproduce the observed features in the shear profile despite the slight deviation of the Cl 0024+17 mass distribution
from axisymmetry.

Kneib et al. (2003) presented a reduced shear profile of Cl 0024+17 at $50\arcsec<r<1000\arcsec$ 
based on two passband (F450W and F814W) WFPC2 observations. In the overlapping ($50\arcsec \lesssim r \lesssim 100 \arcsec$) region, we
find that their shear profile is systematically lower than ours and 
the discrepancy is increasing for decreasing radius. We suspect that the difference mainly comes from the somewhat large dilution of 
the lensing signal from non-background population in their source catalog. With only two passband WFPC2 data available, this
contamination is inevitable. Futhermore, our ACS observations are much deeper
than their WFPC data, allowing us to utilize more distant and more highly
distorted galaxies that were not previously available (i.e., our effective
source plane is at higer redshift). However, it is encouraging to observe that their shear profile
also possesses a similar $dip$ at $r\sim75\arcsec$ (see Figure 7 of their paper).

\subsection{Source Image Reconstruction \label{source_image_reconstruction} }

\begin{figure}
\plotone{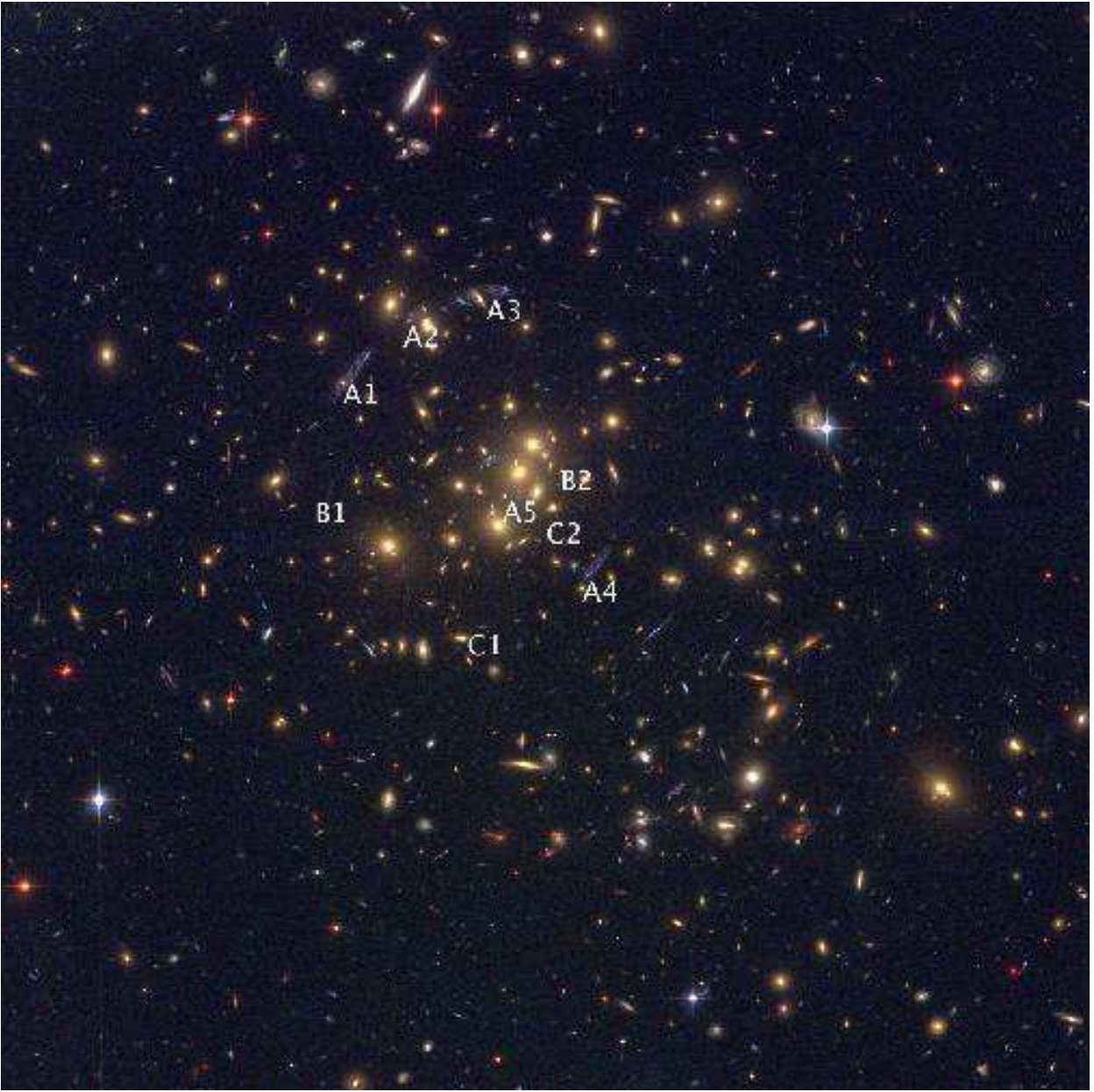}
\caption{HST/ACS color composite of Cl 0024+17 in the observed orientation: North is right and East is up.
The ACS/WFC $g_{475}$, $r_{625}$, and
$z_{850}$ images are used to represent the intensities in blue, green, and red, respectively.
We show the central square ($196\arcsec \times 196\arcsec$) region of the cluster, which
precisely overlaps our mass reconstruction field. The five multiple images of the single
source at $z=1.675$ are labeled as A1-A5. We also denote the two other multiple image system candidates 
(see \textsection\ref{section_implementation}) by B1-B2 and C1-C2.
\label{fig_cl0024}}
\end{figure}

\begin{figure}
\plotone{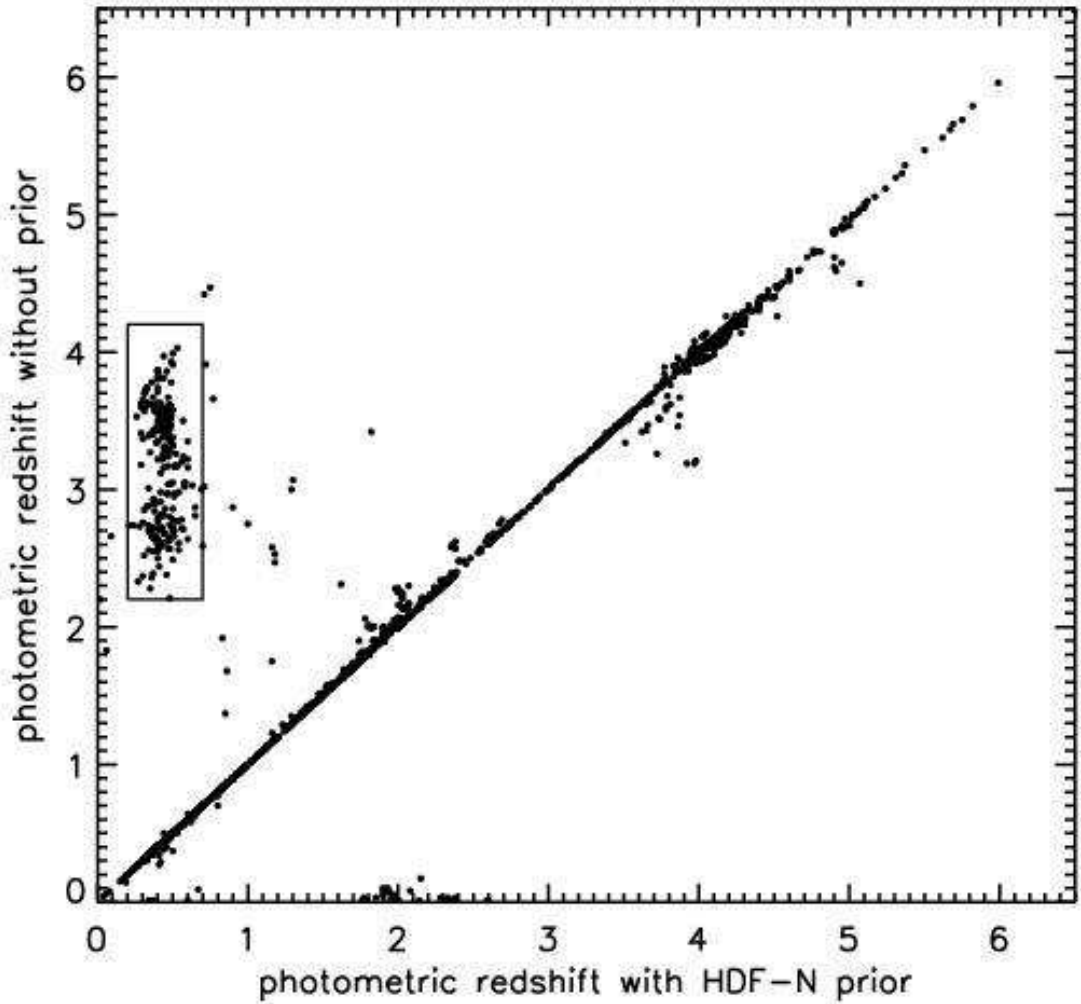} 
\caption{Effects of priors in the photometric redshift estimation. We produce two sets of photometric
redshift catalogs to examine the effect of the HDF-N prior in the presence of lensing.
The comparison shows that no systematic difference
is found between the two sets except for those objects in the box. It appears that these
objects are associated with the main cluster at $z\sim0.4$, but are mistaken for
high-redshift objects when no prior is used.
\label{fig_photo_z_compare}}
\end{figure}

\begin{figure}
\plottwo{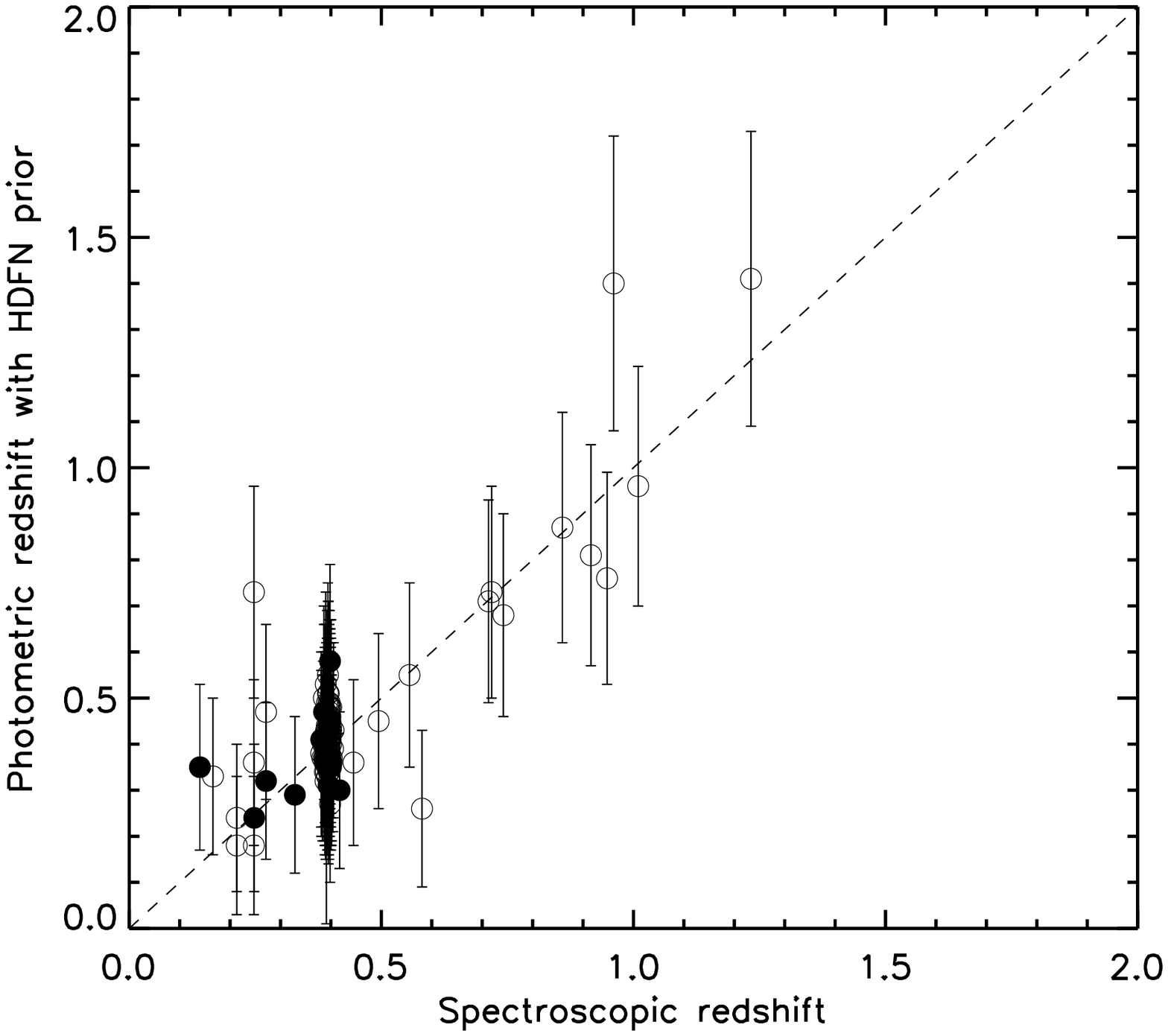}{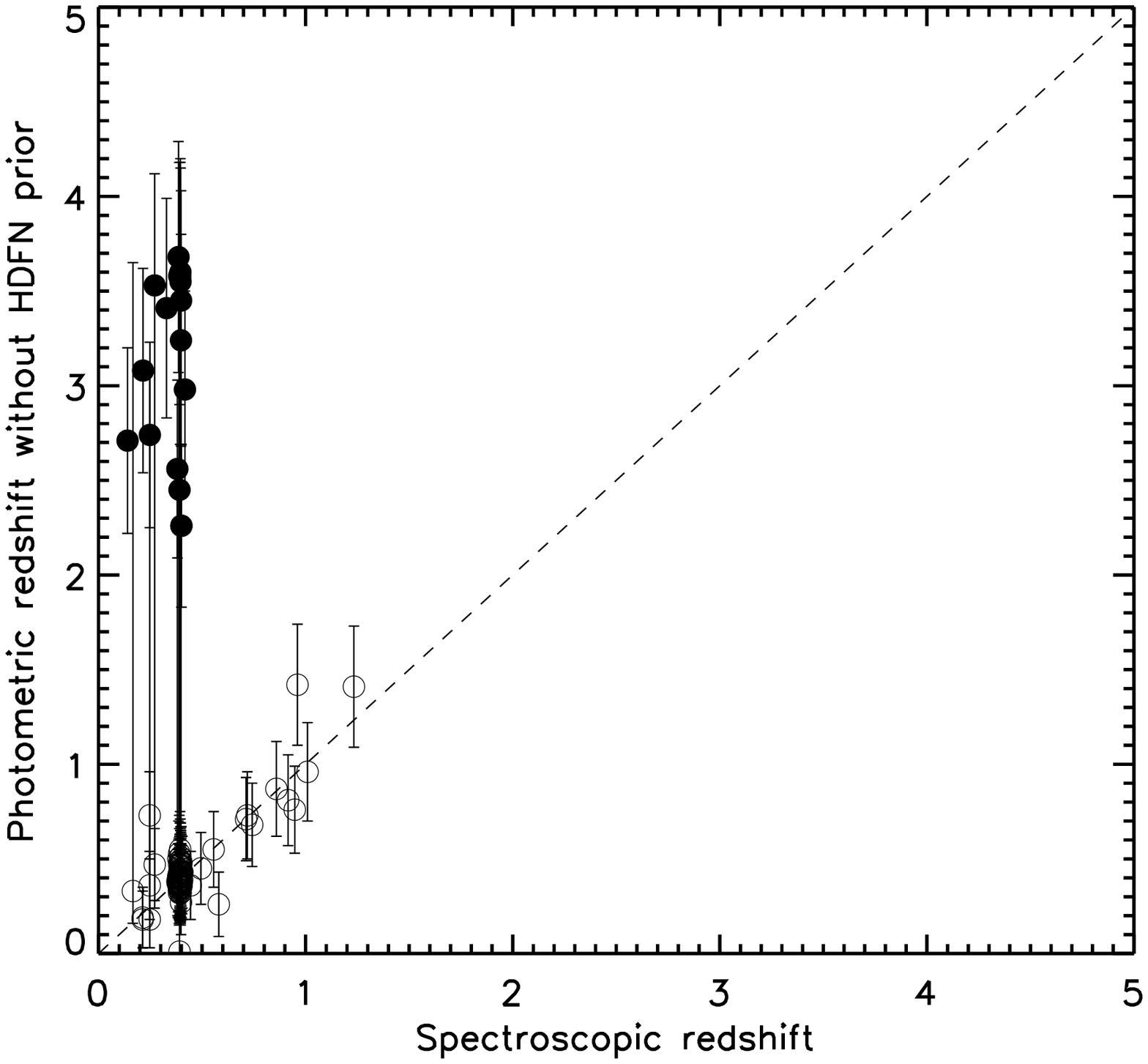} 
\caption{Spectroscopic redshifts vs. photometric redshifts. The photometric redshifts obtained
with HDFN priors are consistent with the spectroscopic redshifts ($left$). The filled
circles correspond to the objects classified as outliers in the right panel. 
When the photometric redshifts are computed without priors, we observe catastrophic
outliers at $z_{phot}=2\sim4$ (see the text for description).
\label{fig_photo_z_spec}}
\end{figure}

\begin{figure}
\plotone{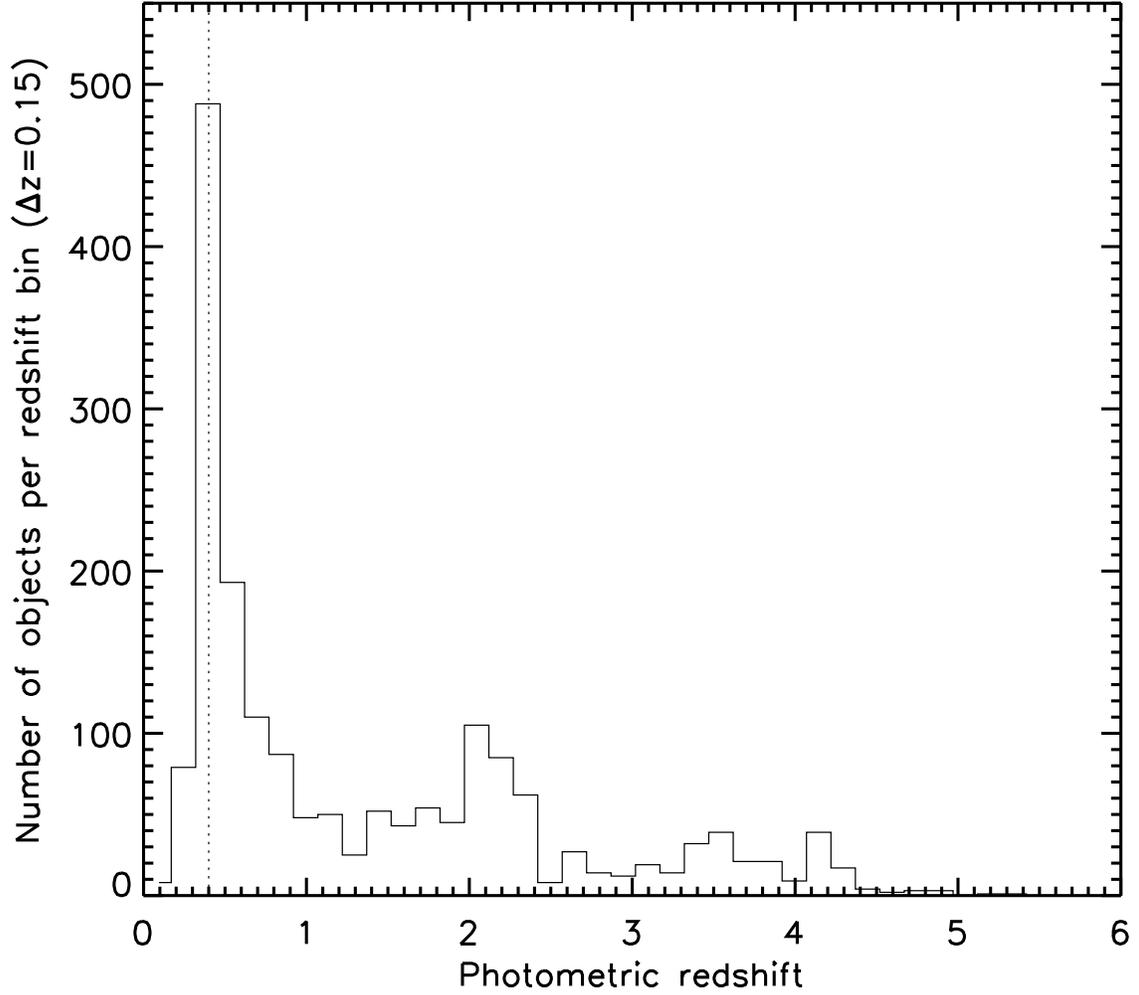} 
\caption{Photometric redshift distribution of the $i_{775}<27.5$ non-stellar objects in the Cl 0024+17 field. The redshifts are estimated
using the six passband ACS photometry with HDFN priors. The redshift spike at $z=0.4$ (dashed line) is clearly visible.
\label{fig_photo_z}}
\end{figure}

\begin{figure}
\plotone{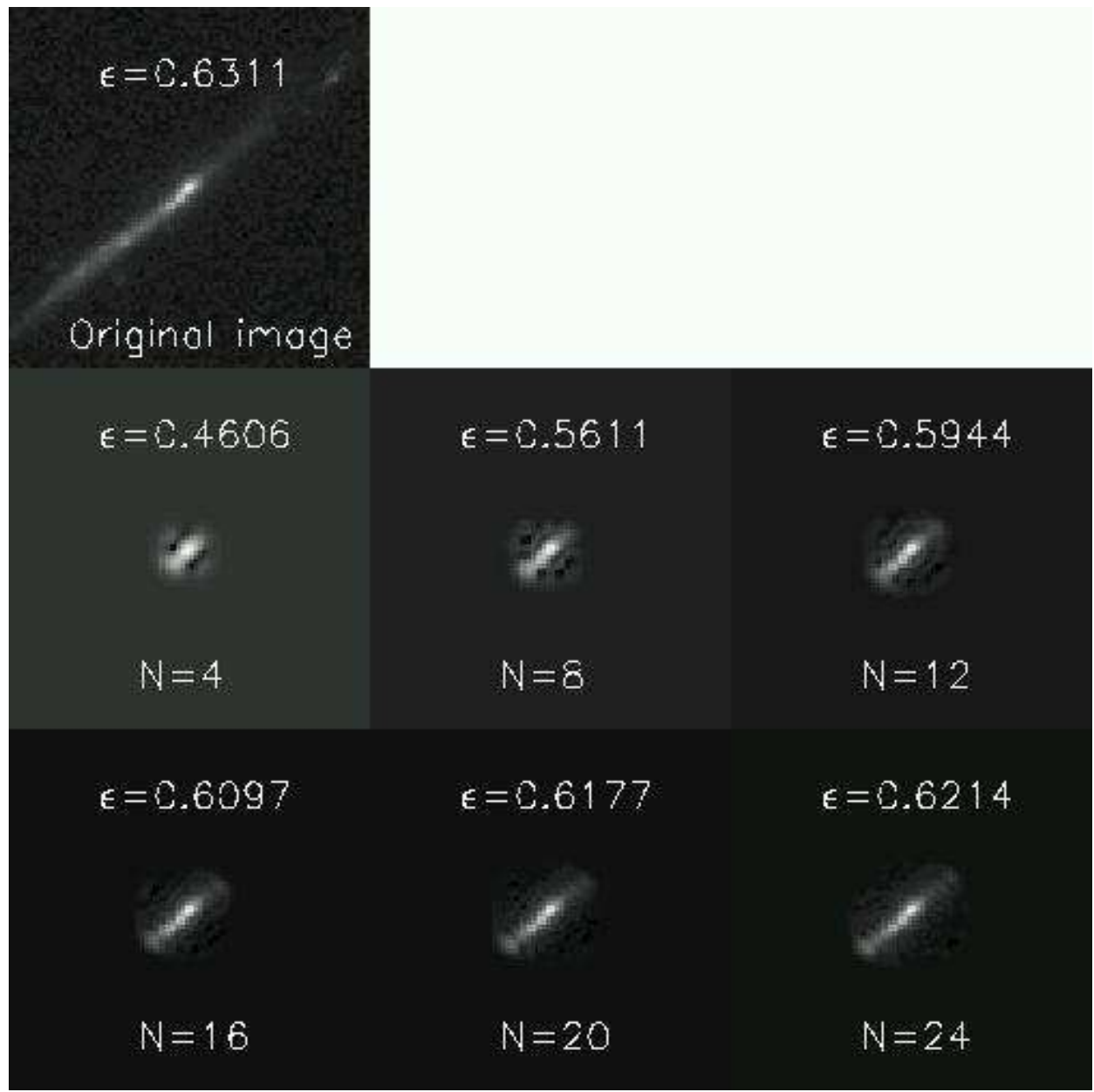} 
\caption{Example of aliasing in a shapelet representation. Intensities are on a square-root scale.
We display the shapelet decomposition and the measured
ellipticity $\epsilon$ of an highly elongated object for different $N$ (shapelet order). When we measure the
ellipticity by directly fitting an ellipticial Gaussian function as proposed in the current paper, 
we obtain $\epsilon=0.6311$. As we increase $N$, the recovered ellipticity from the shapelet method 
approaches this value, yet very slowly.
\label{fig_aliasing}}
\end{figure}

\begin{figure}
\plottwo{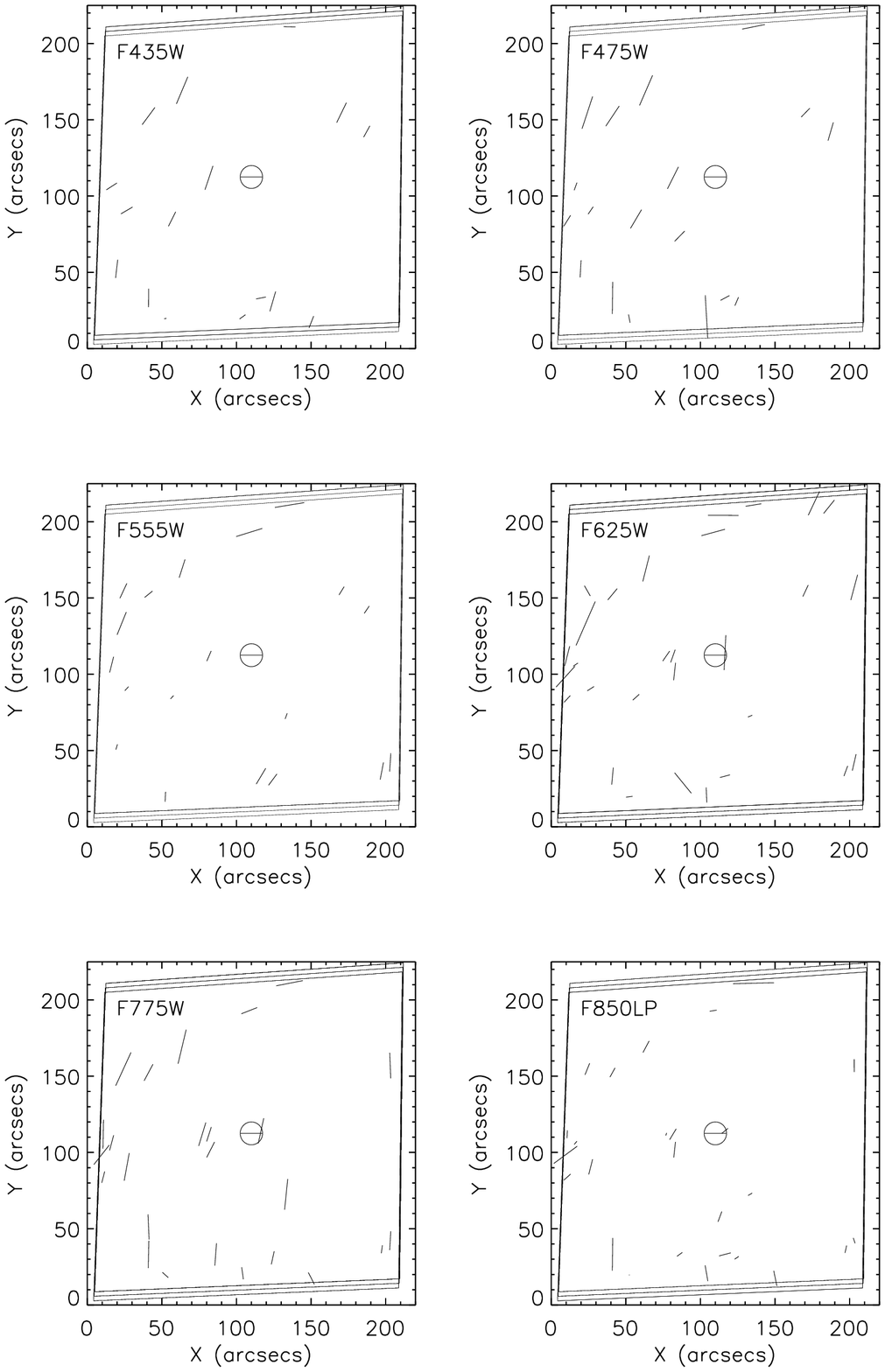}{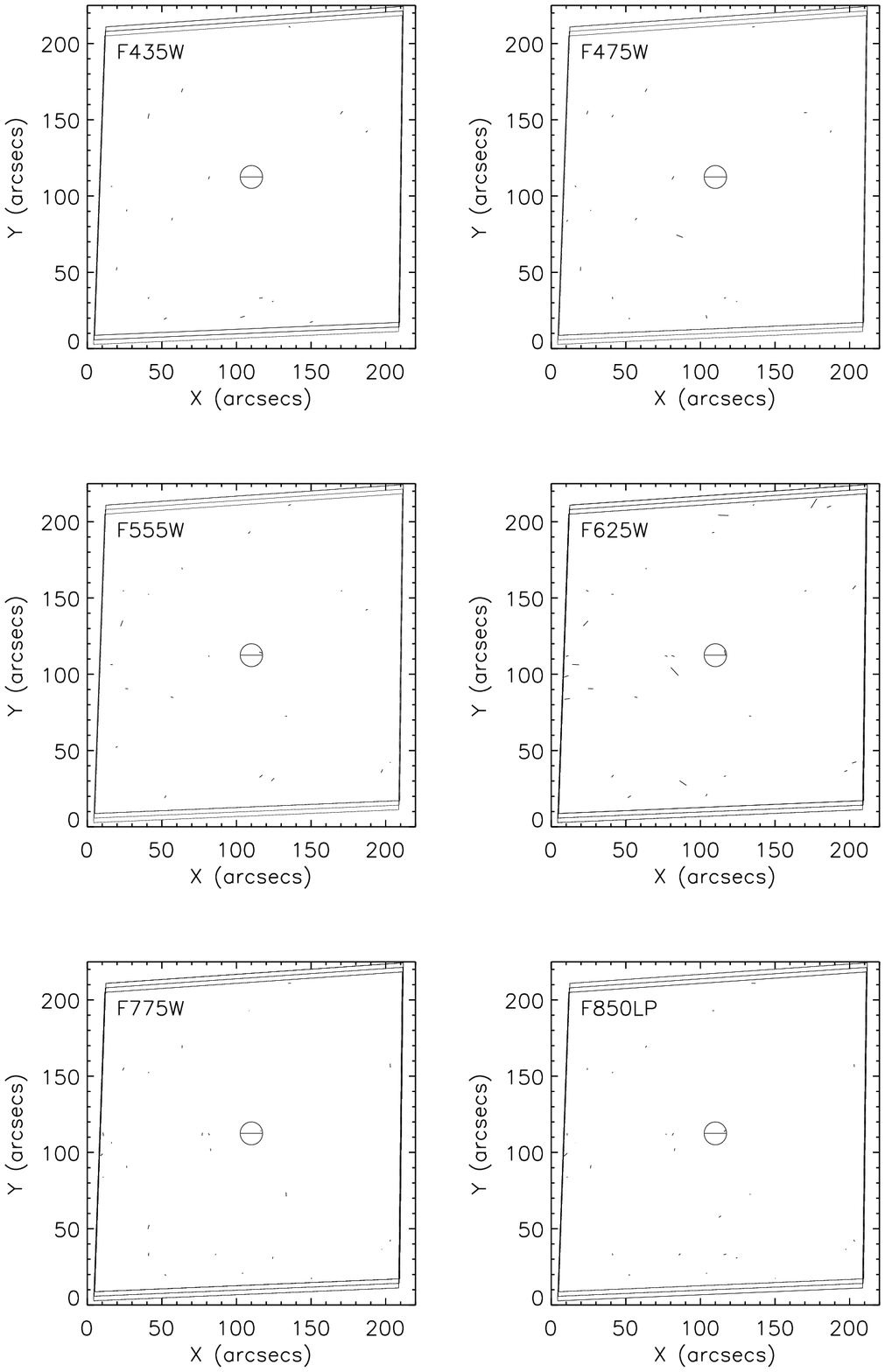} 
\caption{Observed PSF pattern and correction in the ACS/WFC Cl 0024+17 field. 
The stars are selected from the size vs. magnitude plot.
We show the observed
ellipticity pattern of the stars before ($left$) and after ($right$) the correction.
The circled stick in the center illustrates ellipticity of $\epsilon=0.05$.
\label{fig_psf}}
\end{figure}

\begin{figure}
\plottwo{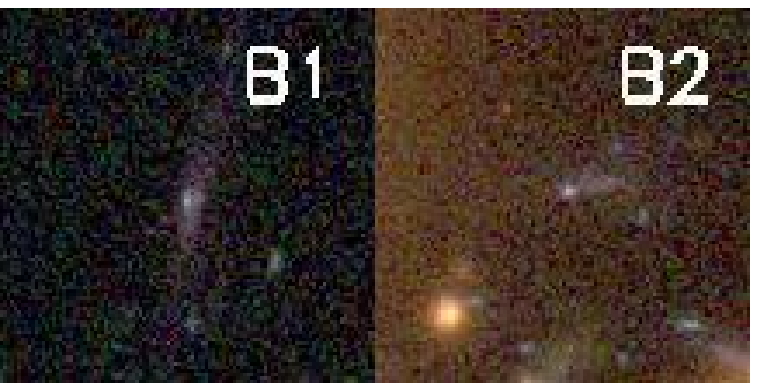}{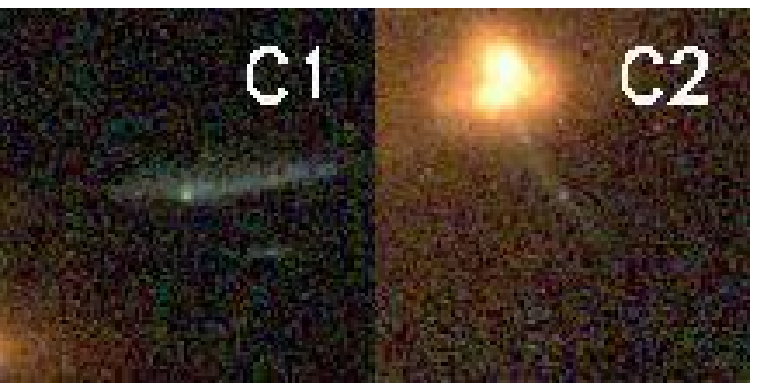} 
\caption{Two additional multiple system candidates used as strong-lensing constraints. The B1-B2
system is originally identified by Broadhurst et al. (2000) with WFPC2 observations. The photometric redshift
of the system is $\bar{z}_{phot}=1.27$. Our initial mass model based on this B1-B2 system along with the A1-A5 system
predicts that the C1-C2 images with $\bar{z}_{phot}=2.84$ are also multiply lensed.
\label{fig_new_multi}}
\end{figure}

\begin{figure}
\plotone{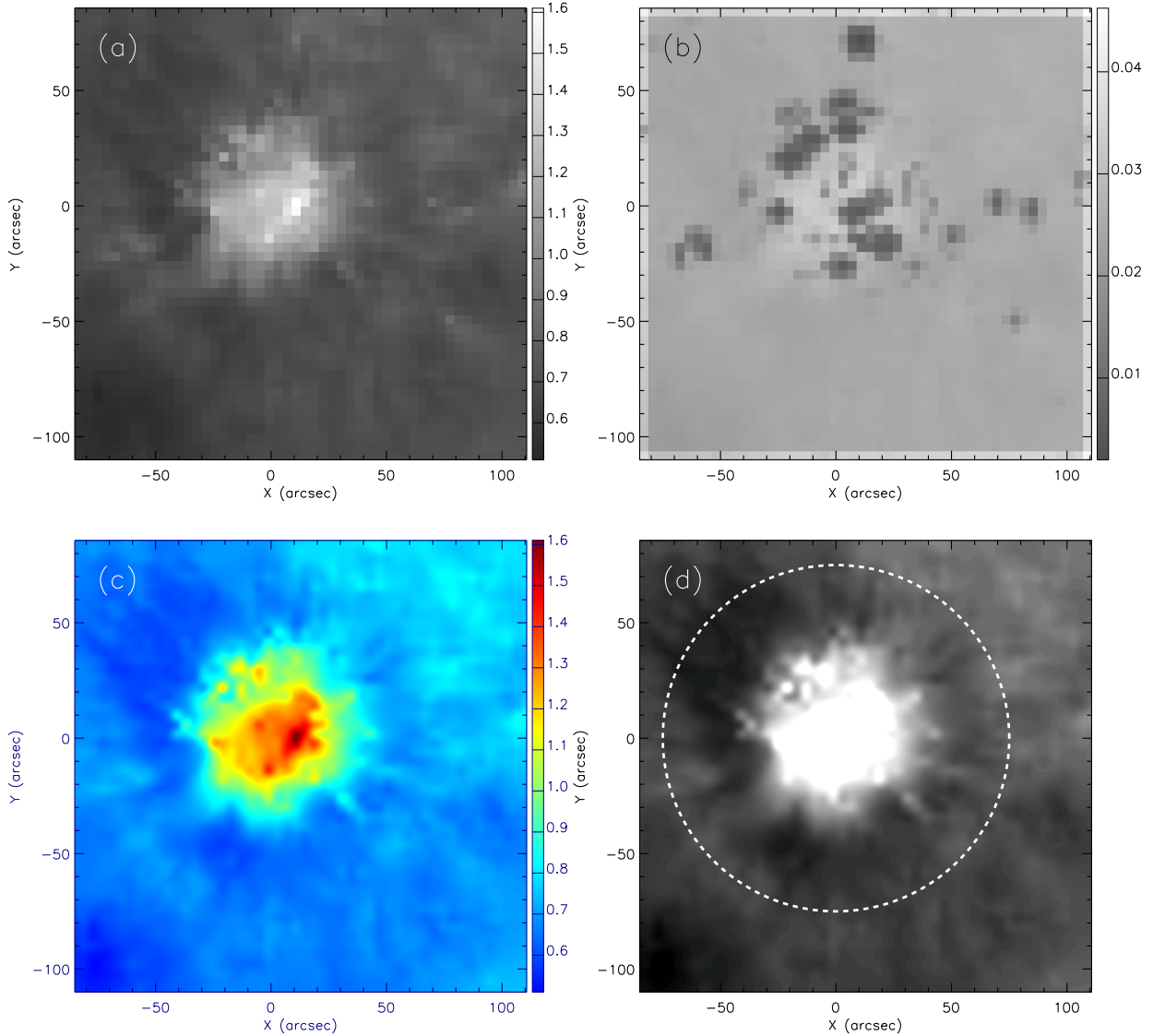} 
\caption{Mass reconstruction of Cl 0024+17. 
(a) The $50\times50$ mass map derived from the converged $52\times52$ deflection potential.
(b) The uncertainty of the mass reconstruction derived from the Gaussian approximation (\textsection\ref{section_uncertainty}).
(c) Same as (a), but we reproduced the map with a slightly larger regularization constraint.
The bicubic interpolated result is displayed with a stretched color table to emphasize the 
low-density feature. The color version of this figure is available on-line.
(d) Same as (c), with a dashed circle overlaid at $r=75\arcsec$ to trace the ringlike substructure observed 
at that radius.  We choose the origin to be the geometric center of the ringlike structure.
\label{fig_massmap}}
\end{figure}

\begin{figure}
\begin{center}
\plotone{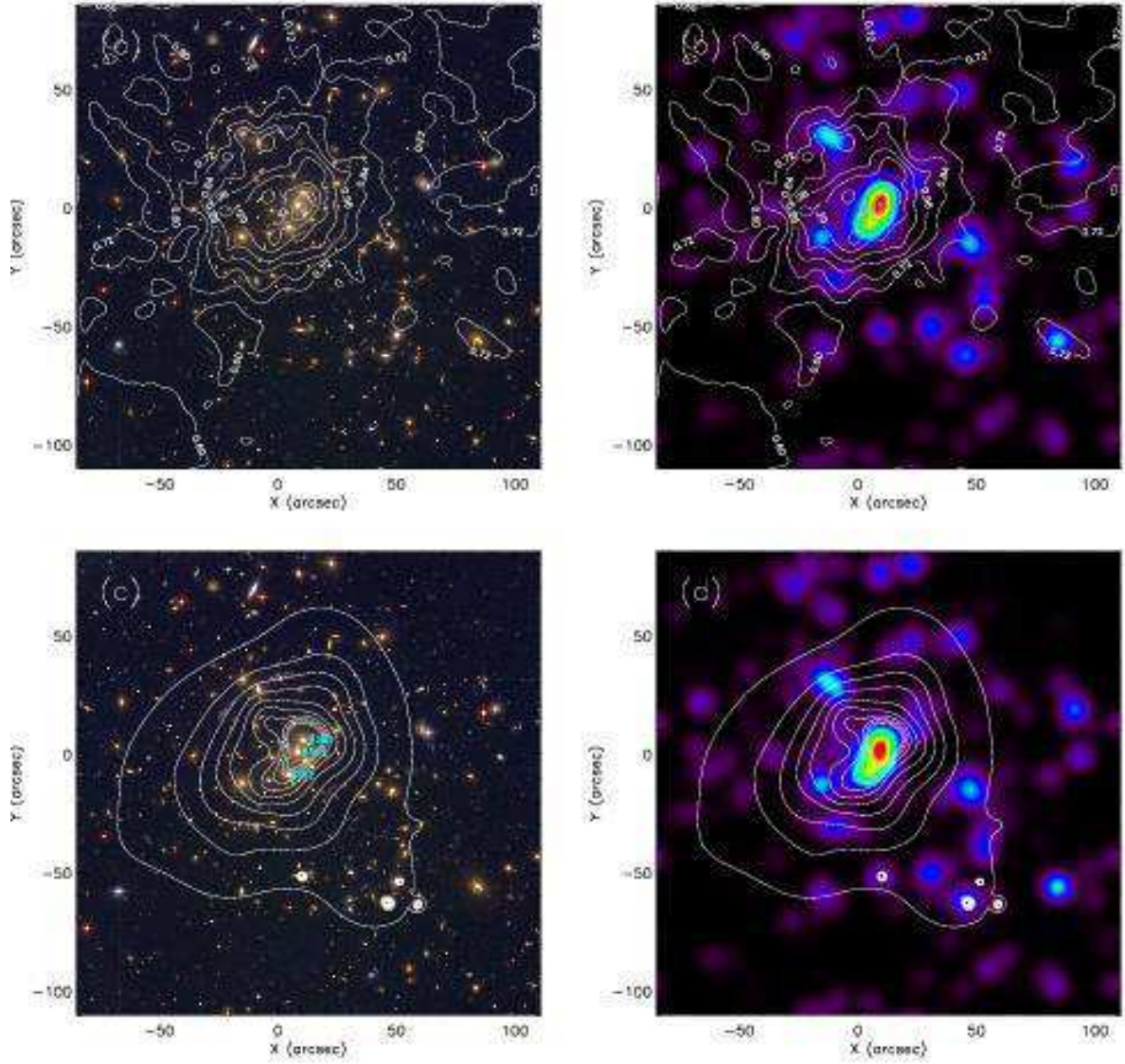}
\end{center}
\caption{Cl 0024+17 mass and X-ray overlaid on cluster lights. (a) Mass contours overlaid on the
ACS color composite.
(b) Mass contours on the smoothed (FWHM$\sim10\arcsec$) cluster light ($r_{625}$) distribution. .
(c) Chandra X-ray contours on the ACS color composite. The X-ray image was exposure-corrected and
adaptively smoothed (Ebeling et al. 2006) with a mininum significance of $3~\sigma$. 
(d) Same as (c), but the background is replaced with the smoothed
light distribution.
\label{fig_xraymap}}
\end{figure}

\begin{figure}
\plotone{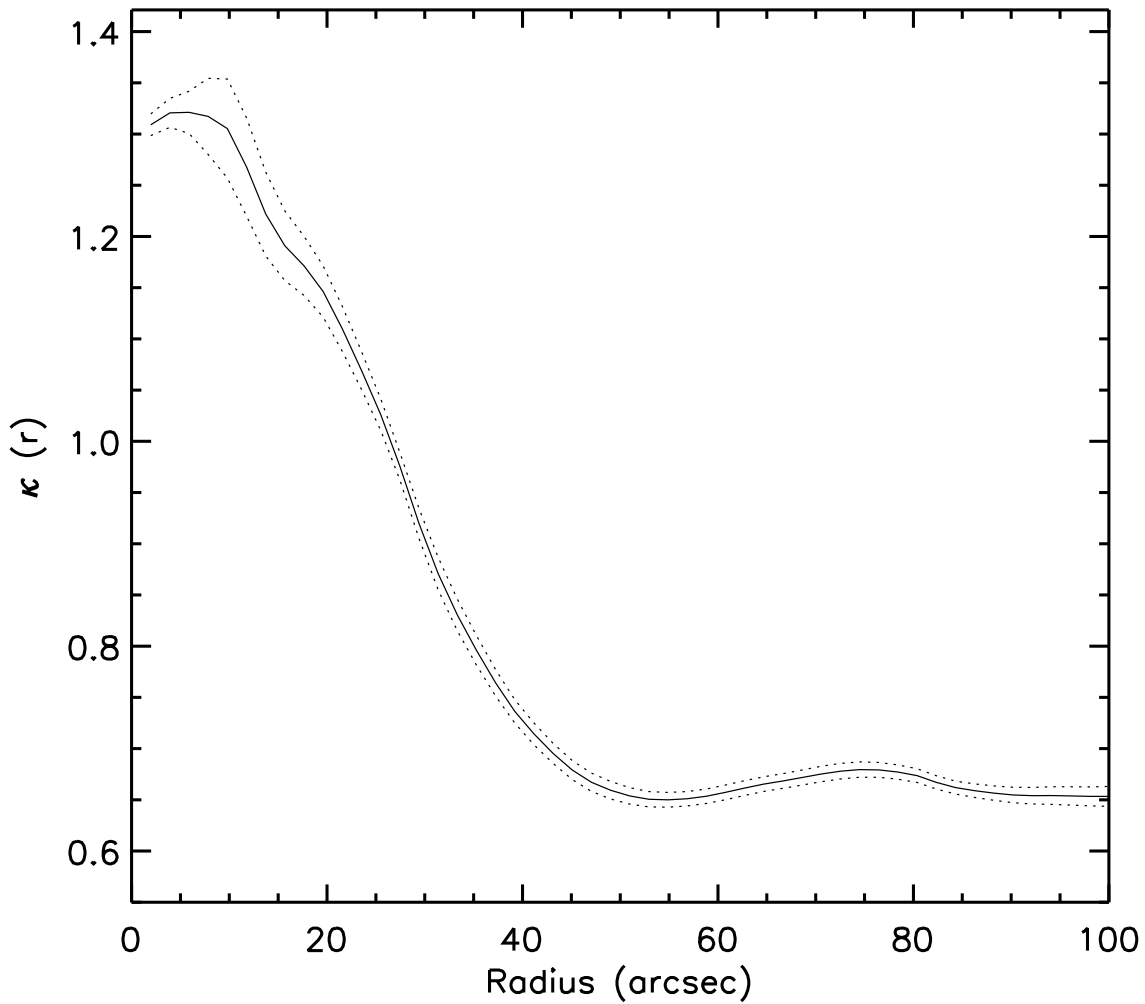} 
\caption{Radial mass density profile of Cl 0024+17. The mass density ($solid$) is given in units of critical density $\Sigma_c$ at a fiducial
redshift of $z_f=3$. The dotted lines represent the 1 $\sigma$ deviation of the azimuthal average. 
It is clear that the radial mass profile
of the cluster is peculiar and does not resemble any conventional analytic profile. At $r\lesssim50\arcsec$, $\kappa$
decreases for increasing $r$. Outside the $core$ ($r\gtrsim50\arcsec$), $\kappa$ rises and creates a ``bump'' at $r\sim75\arcsec$.
\label{fig_radial_density}}
\end{figure}

\begin{figure}
\plotone{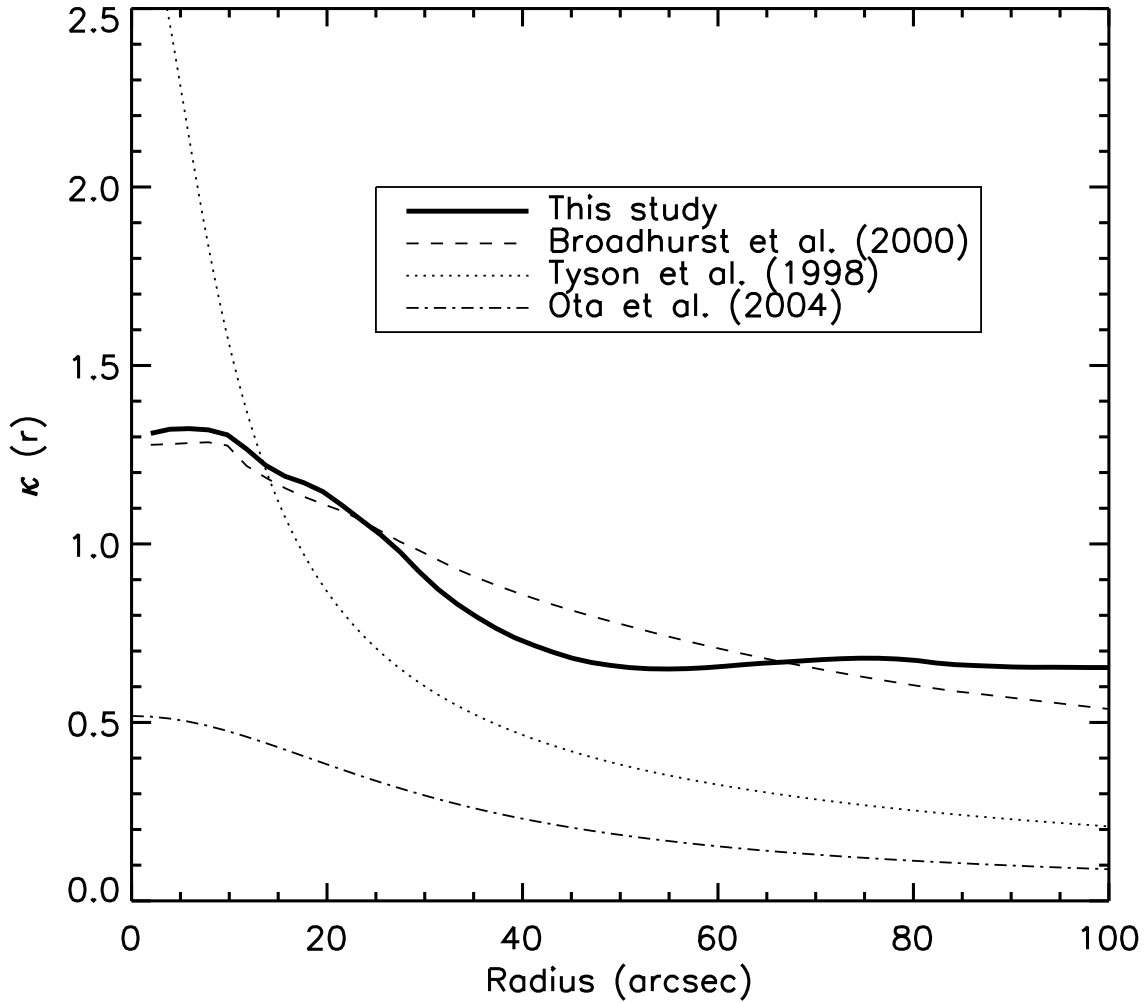}
\caption{Density profiles of Cl 0024+17 from different studies. The overall shape of our radial density profile ($solid$) 
looks more striking when compared
to the results of the previous studies. 
We transformed the results of Tyson et al. (1998) ($dotted$), Broadhurst et al. (2000) ($dashed$),
and Ota et al. (2004) ($dot~dashed$) using the current cosmological parameters. Note that Tyson et al. (1998) and
Broadhurst et al. (2000) derived the mass density using strong-lensing while Ota et al. (2004) used $Chandra$
X-ray observations.
As already indicated by Ota et al. (2004),
the X-ray mass is far less than the other three lensing results; a more recent X-ray
analysis with XMM-Newton (Zhang et al. 2005) (omitted here) yields even slightly lower values. The low core densities 
($\kappa < 1$) predicted by these X-ray analyses violate
the fundamental condition of the strong-lensing, which requires a projected mass
density greater than unity. 
\label{fig_density_compare}}
\end{figure}

\begin{figure}
\plotone{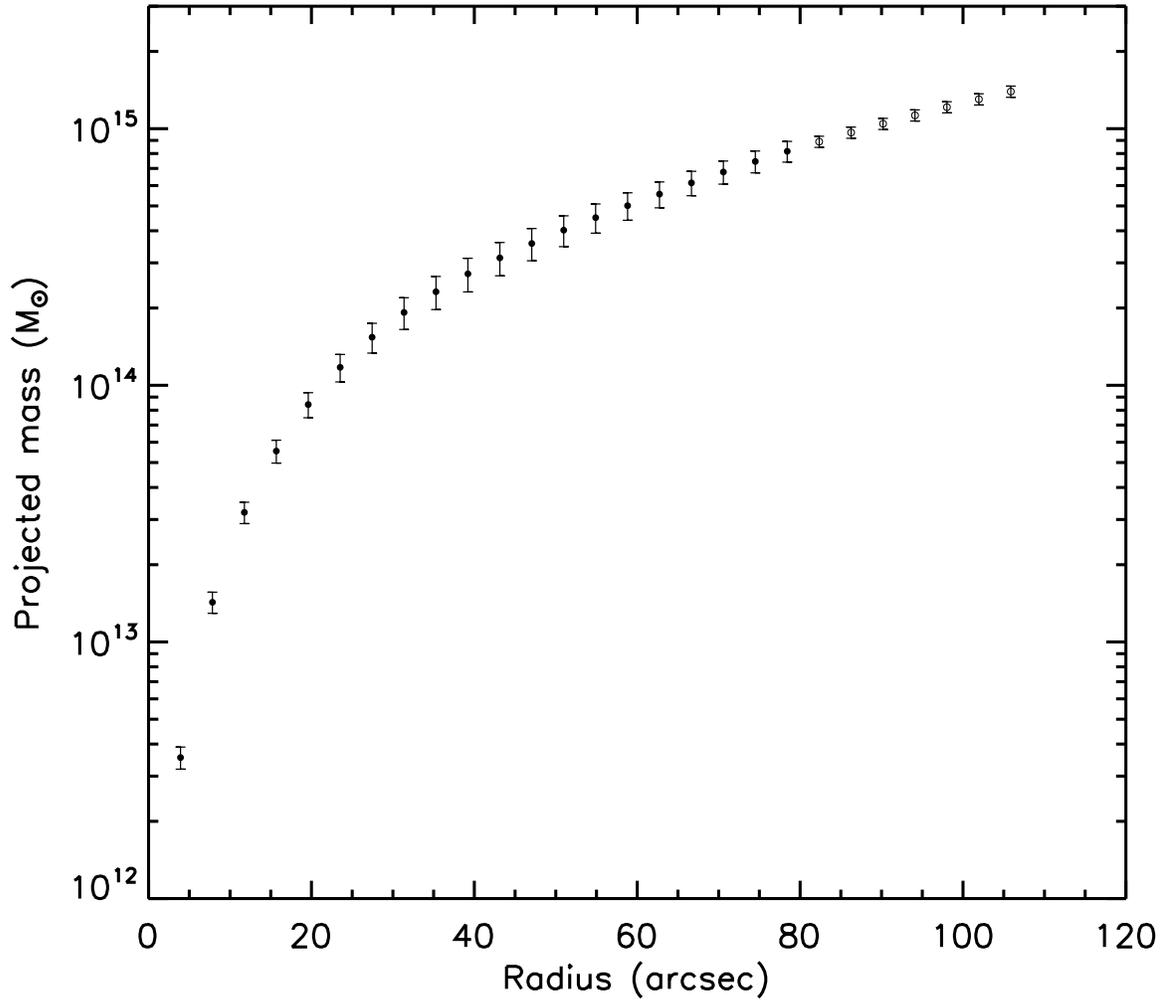}
\caption{Projected mass in the cluster core. The mass is directly computed from
the mass map in Figure~\ref{fig_massmap} by adding the mass pixel values within
a given aperture (filled circle). At $r\gtrsim80\arcsec$ we cannot complete a circle within the
field and thus we assume an axisymmetry to extend the profile out to $r\sim110\arcsec$ (open circle).
\label{fig_projected_mass}}
\end{figure}

\begin{figure}
\begin{center}
\plotone{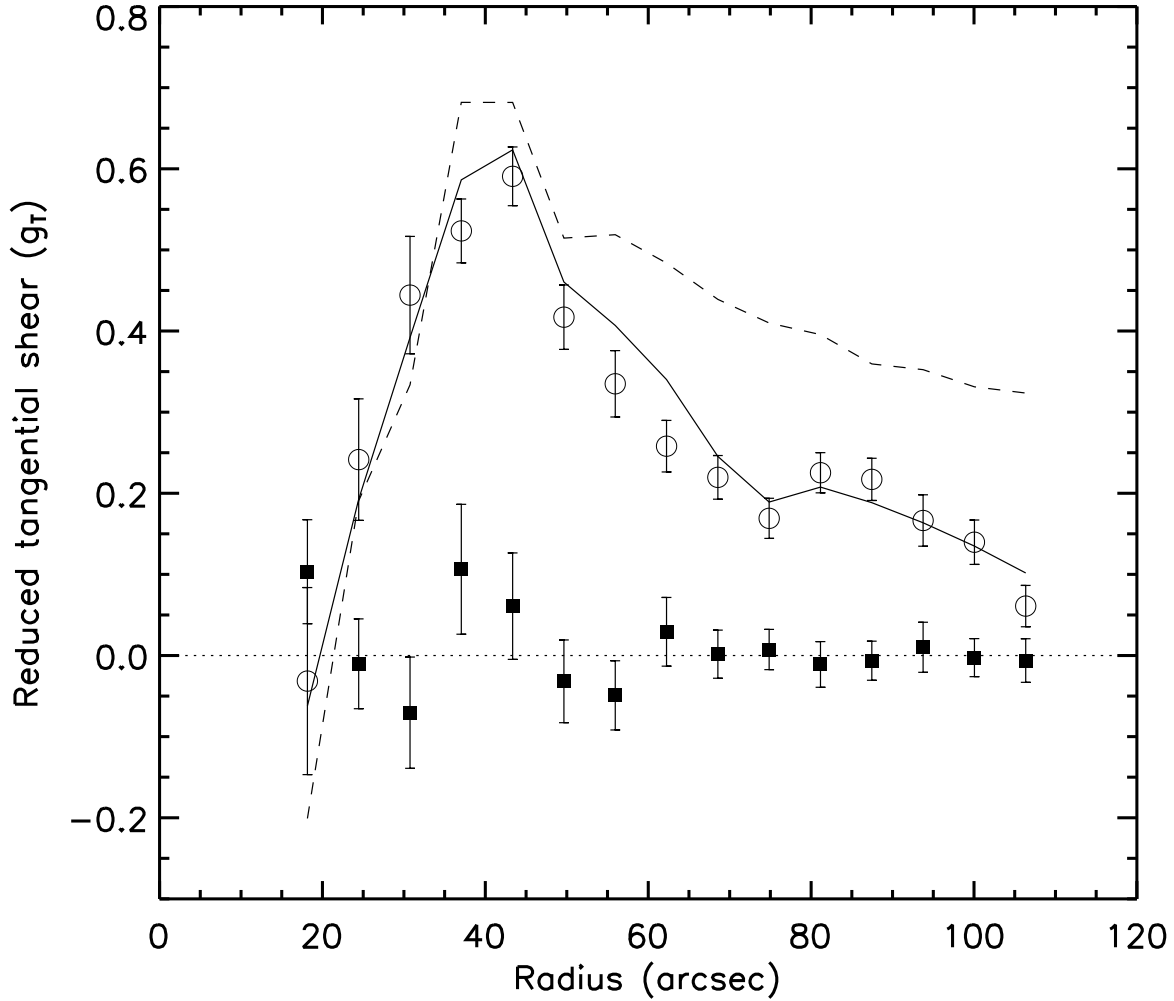}
\end{center}
\caption{Reduced tangential shears of Cl 0024+17. Open circles represent the reduced
tangential shears measured from $\sim1300$ background galaxies. 
We note that there is a dip at $r\sim75\arcsec$, which indicates the presence of the ringlike
sub-structure seen in the 2-D mass reconstruction or the ``bump'' in the radial density profile.
We display the predicted shears estimated from our mass profile ($solid$).
Also shown is the predicted tangential shears when the Broadhurst et al. (2000) model is assumed ($dashed$).
We also performed the 45\degr ~rotation (B-mode) test to examine possible
systematics. As observed ({\it filled square}), the lensing signal disappears in
this case and the residual amplitudes are consistent with the statistical errors.
\label{fig_tan_shear}}
\end{figure}

\begin{figure}
\plotone{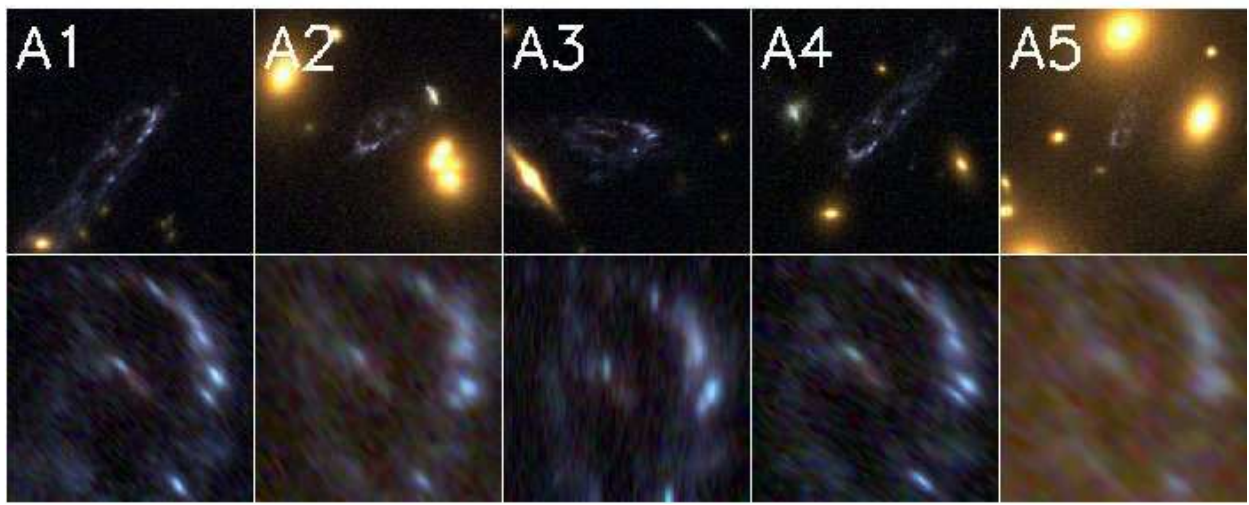}
\caption{Source image reconstruction of the five well-known multiple images at $z=1.675$.
The top row shows the observed lensed images directly cut from Figure~\ref{fig_cl0024}.
In the bottom panel, we display the delensed image of each arc predicted from our deflection potential.
Note that the orientation, parity, and size of these delensed images are consistent with
each other. The size of the delensed source images are approximately $0.4\arcsec\times0.5\arcsec$.
\label{fig_source_reconstruction}}
\end{figure}

The ability to correctly reproduce the observed multiple images is a necessary but not
sufficient condition for a robust mass model, especially if the strong-lensing data are sparse.
To put our model to the test, we regridded the $50\times50$ deflection field 
into the $3920\times3920$ grid (matching the resolution of the ACS/WFC cluster image
shown in Figure~\ref{fig_cl0024}) using bicubic interpolation and
performed delensing of the well-known five multiple images at $z=1.675$.
The top panel of Figure~\ref{fig_source_reconstruction} shows the observed lensed
images directly cut from Figure~\ref{fig_cl0024}. In the bottom panel, we display
the delensed image of each arc predicted from our deflection potential.
It is apparent that the orientation, parity, and size of these delensed images are highly consistent
with each other. Colley et al. (1996) presented the first source delensing of the three well-resolved arcs 
(corresponding A1, A3, and A4 in our nomenclature)
from the WFPC2 image analysis. Their reconstructed images have similar orientations (note that
north is up in their image) and ellipticities to ours with identical parity. Nevertheless, the high sampling resolution
of ACS and the improved mass modeling allow us to obtain the delensed images in greater detail.

We also examined the result alternatively by relensing one of these source images back 
to the image plane. Initially, the result was less than ideal, yielding more than the five known multiple
images. The relensed images at the location of the five known arcs were in good agreement with the observation whereas
the rest of the predicted images at other locations look much less definite. Nevertheless, this result should not 
discredit the mass model because noise can cause the deflection field at any arbitrary location to coincidently point to
the same location. We were able to easily fix the problem manually by slightly perturbing the deflection potential of the region 
where {\it false} images were predicted. The resulting mass map looks virtually identical to the original one and
the minimizing function $f$ (eqn.~\ref{eqn_minimization}) is as small as the initial value.

Because only their photometric redshifts are available with less distinctive morphology,
we imposed only weak constraints on the two other multiple systems at $z_{phot}=1.3$ and $2.8$
(\textsection\ref{section_implementation}).
The delensed positions of B1-B2 at $z_{phot}=1.3$ agree nicely and their images in source plane
are similar to each other. We note that the source positions of C1-C2 are also close, yet slightly separated
by $\sim40$ pixels ($\sim2 \arcsec$); this makes the $\chi^2_{\mu}$ values per constraint
rather large for C1-C2 ($\sim3$) whereas they remain close to unity for A1-A5 and B1-B2.

When we forced the two locations to coincide in our
mass reconstruction, the smoothness of the resulting mass map was compromised. Since 
the spectroscopic redshifts of the source is unknown, the solution obtained in this way
cannot be claimed to be better and thus we chose to accept the original result 
as our final mass model.

\subsection{Other Sources of Errors and Their Impacts}

It is certain that we did not consider all sources of errors in our error analysis. Important among these are
photometric redshift uncertainties and the large-scale structures in the ACS field of Cl 0024+17.

Because lensing signals depend on all masses along the line-of-sight between the observer and sources,
some large-scale structures in front of and behind the cluster can contribute to the lensing signals.
In Jee et al. (2005b), we integrated the power spectrum from us to the effective source plane in order
to estimate the contribution in the evaluation of the total mass of the high-redshift cluster 
MS1054-0321 at $z=0.83$. We found that the error introduced by this cosmic shear amounted to $\sim$14\% of the total
cluster mass within 1 Mpc. Therefore, it was a significant factor in the total error budget for that cluster.
Since the similar formalism to compute the cosmic shear contamination has not been developed for the current mass
reconstruction method, we do not attempt to estimate the correponding errors for Cl 0024+17. Nevertheless, we
suspect that the cosmic shear induced error is substantially smaller for the current study because 
the lens is at much lower redshift where the cluster lensing efficiency is much higher. 

The typical uncertainty of $\delta z\sim0.1$ in photometric redshift estimation does not greatly affect the evaluation
of the expected reduced shear in Equation~\ref{eqn_mean_ellipticity}. The cosmological weight function
(eqn.~\ref{eqn_cosmo_weight}) changes slowly with source redshift when the lens is at $z=0.4$. Hence, the change
in the expected ellipticity due to the uncertainty of $\delta z\sim0.1$ is much smaller than the intrinsic
ellipticity dispersion in many cases. 

What will happen to the cluster mass profile if the redshifts of the B1-B2 and C1-C2 systems that we used as strong-lensing 
constraints have significantly large photometric redshift errors ($\delta z>0.1$) and correct spectroscopic redshifts become 
available in the future?
Because B1-B2 and C1-C2 are used to lift the mass-sheet degeneracy, the new mass map will be 
a simple invariant [$\kappa \rightarrow \lambda \kappa + (1-\lambda)$] transformation of the current mass map.
Therefore, the overall shape of the mass profile and the significance of the ring feature 
at $r\sim75\arcsec$ will not be greatly affected. We also repeat that the transformation will not change the total
projected mass within the Einstein radius.

Finally, despite the deep, six-passband ACS photometry and the use of HDFN priors, we expect that there still
is a small fraction of non-background population in our source catalog due to some catastrophic errors
in photometric redshift estimation.
Because of the small ACS field and the lack of known spectroscopic sample for faint galaxies ($i_{775} > 24$),
it is difficult to estimate this fraction reliably. As we did not account for the dilution of the lensing
signal from the contamination, one may argue that our mass estimates are lower limits. However, in the
nonlinear lensing regime where we combine strong- and weak-lensing constraints, 
the effect is reversed. For example, if the contamination is removed somehow
in our source catalog, the reduced shear profile (Figure~\ref{fig_tan_shear}) will be shifted upward. The higher shear
profile certainly implies a steeper mass profile in the nonlinear regime. Therefore, the cluster mass decreases more rapidly as
$r$ increases in this case.

\section{DISCUSSION \label{section_discussion}}
\subsection{ringlike Structure as Direct Evidence of a Line-of-Sight Collision \label{section_nbody}}
The first observational indication that Cl 0024+17 might have undergone a high-speed line-of-sight collision
was presented by Czoske et al. (2002) based on their wide-field spectroscopic survey of the cluster.
Their redshift histogram obtained from the $\sim300$ spectroscopically confirmed cluster members
shows that the redshift distribution of the cluster is bimodal; the larger peak is
at $\bar{z}=0.395$ and the smaller peak at $\bar{z}=0.381$.
This bimodality becomes more distinct if the histogram is reproduced only using the cluster galaxies 
at $r>200\arcsec$. On the other hand, when the redshift distribution at $r<200\arcsec$ is
examined, the separation between the two peaks is not clear; it rather appears that
there is one main clump at $\bar{z}=0.395$ yet skewed towards negative velocities.
This unusual redshift distribution of the cluster led Czoske et al. (2002) to suggest a scenario wherein
the system underwent a high-speed line-of-sight collision of two sub-clusters with a mass ratio of 2:1 a few
Gyr ago, and
the negative velocity tail and the smaller peak originally had belonged to the less massive system.
They also supported their scenario with an $N$-body simulation, which predicts the
observed velocity distribution.

We argue that the ringlike dark matter sub-structure and the flat density profile of Cl 0024+17
in our high-resolution mass reconstruction provide alternative yet much stronger evidence
for the line-of-sight collision and can be used to refine the scenario of Czoske et al. (2002).
A high-speed collision of two massive clusters can be approximated by a gravity impulse
at the cluster center lasting $\Delta t$, whose order
is the size of the cluster divided by the impact velocity.
Because of an increased gravity, the two clusters contract for the duration of the impulse $\Delta t$. 
When the impulse is over,
the contraction stops and both clusters start to expand.  The extra kinetic energy
causes the dark matter in the cluster outer regions to become unbound and to scatter to a large distance.
The dark matter in the cluster inner regions will also expand. However, because it is still bound to the clusters,
the expansion slows down. This deceleration leads to a crowding of orbits or a shell, which should appear
as a ringlike structure around the cluster cores in projection. 

One challenging question in the interpretation of the ripple-like structure
is, however, how efficient the gravitational shock is on a cluster scale. The aforementioned ring creation mechanism
is analogous to that in ring galaxies (Lynds \& Toomre 1976). Quite a few numerical simulations have shown that the ringlike
structure can arise from a radial density propagation in a high-speed collision of two galaxies (e.g., Hernquist \& Weil 1993).
Nevertheless, we need to examine if the argument for ring galaxies can also apply to a collision
of two spheroids on a cluster scale.

To investigate the problem, we perform a purely collisionless $N$-body simulation of a collision of
two massive clusters similar to the Czoske et al. (2002) experiment. The mass ratio is set to
2:1 and both clusters follow a softened isothermal distribution (i.e., $\rho \propto (1 + (r/r_c)^2)^{-2}$).
The core radius of the larger cluster is 100 kpc and has a mass of $6\times10^{14} M_{\sun}$
within a 2 Mpc radius whereas the core radius is chosen to be 60 kpc for the smaller cluster with
$3\times10^{14} M_{\sun}$ within a 1 Mpc radius. The number of particles for the larger and smaller
clusters is $2\times10^5$ and $10^5$, respectively; thus a particle mass is $3\times10^9 M_{\sun}$.
The initial separation is 3 Mpc with a relative velocity of 3000 $\mbox{km}~\mbox{s}^{-1}$.
A Plummer force softening length is set to 5 kpc.
Note that although our choice of these parameters may
resemble the hypothesized two clusters of Cl 0024+17 before the collision, we do not elaborate to
refine them to ensure the desired final result. Our main goal here is to examine whether the
observed density structure can occur in a purely collisionless encounter of two massive clusters.
The simulation was carried out with the publicly available GADGET-2 software (Springel 2005) in a
Newtonian space. The forces were computed through the tree algorithm.

\begin{figure}
\begin{center}
\plotone{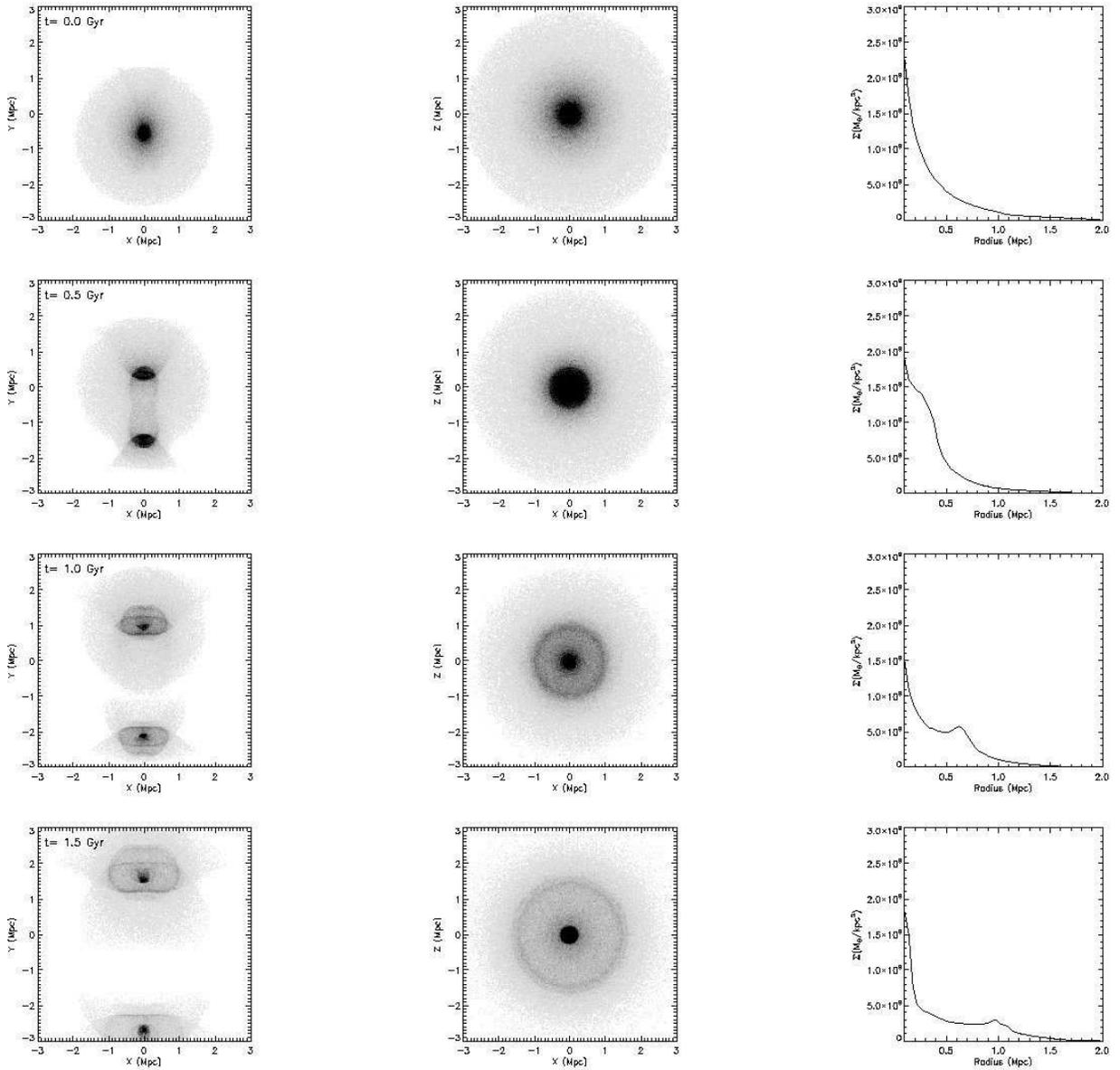}
\end{center}
\caption{Numerical simulation of two colliding clusters. The mass ratio is set to 2:1 and both
clusters follow a softened isothermal distribution (see text for parameters). 
Each row shows snapshots
of the collisionless $N$-body simulation at a given epoch (t is a elapsed time since the core impact). Particle distribution
is projected onto the $x-y$ plane ($left$; the plane containing the collision axis) and the $x-z$ plane ($middle$;
viewed along the collision axis). We also illustrate the projected ($x-z$ plane) density
profile in the right column. A radially expanding shell is visible in the shapshots $\sim 1$ Gyr after the core impact, which also
produces a prominent ``bump'' in the radial mass profile.
\label{fig_nbody}}
\end{figure}

We present four snapshots of the $N$-body simulation in Figure~\ref{fig_nbody} at 0.5 Gyr intervals from the $t=0$ impact moment 
to $t=1.5$ Gyr after the core pass-through. In the $t=0.5$ Gyr snapshot,
the cores of both clusters start to expand and the radial density profile plot shows the resulting
disruption. We can also observe the cluster outer regions start to stream radially.
At $t=1$ Gyr, the two cluster cores are separated by $\sim3$ Mpc and it is clear that
the slowing-down of the expanding particles causes the formation of shell-like structures around both cores.
The shells are rather flattened perpendicular to the collision axis and appear as ringlike
structures when projected along the collision axis. The radial profile also shows a corresponding peak
at $r=0.6$ Mpc. About 1.5 Gyr after the core passage, the shell-like structures are still present and
expanding. We observe that these features last even after a few Gyr.
By iterating the above simulation with different initial conditions (e.g., replacing the
softened isothermal halos with cuspy profiles),
we verify that
these qualitative mass structures are somewhat ubiquitous in high-speed collisions although
the details differ. We stress that the shells also arise for moderately off-center collisions.
The ringlike structure and the small bump in the radial density profile seen in Figure~\ref{fig_nbody}
resemble the two-dimensional mass map (Figure~\ref{fig_massmap}) and the radial mass profile (Figure~\ref{fig_radial_density})
of Cl 0024+17, respectively. Many factors determine
the radius of the ring in the simulation as a function of the elapsed time, including mass ratios, core
radii, impact velocities, etc. Nevertheless, we speculate from the dissipation of the shock in the X-ray
observation and the size of the observed ring in
our mass map that the two clusters in Cl 0024+17 collided perhaps $1-2$ Gyr ago. 

The exact representation of the mass structure of Cl 0024+17 requires
fine-tuning of the initial conditions with the inclusion of the cluster ICM and will be
a subject of future investigations.
The numerical simulation of cluster mergers by Ricker \& Sarazin (2001)
included both dark matter and gas particles.
Their head-on merger simulation with a 1:3 mass ratio demonstrates
that both the dark matter and the gas components of the clusters survive the core-passage and reach their maximum
separation in a timescale of sound crossing time ($\sim1.9$ Gyr) though the gas components suffer severe
distortions and thus are slightly displaced from the corresponding dark matter halos.
Since their analyses were focused on the global X-ray properties of the merging clusters, the time evolution of the detailed dark
matter
profile was not investigated. Nevertheless, it appears that shell-like features are absent in their
snapshots of the merger simulation. We suspect that the employed particle-mesh (PM) force computation did
not provide the resolution and may have smoothed out small scale features.

Having discussed the possible scenario for the formation of the ringlike structure with the numerical simulation above,
we now consider two issues relating to the observational features of the ringlike structure.
First, we note that there are azimuthal variations in the observed ``ring''. The feature appears
to be strongest in the lower-left (southwest) quadrant and weakest in 
the upper-left (southeast) corner. Obviously, the mass distribution in real
clusters are not symmetric. Hence, the ring arising from the collision
should reflect somehow the previous asymmetry. In addition, as already mentioned above,
moderately off-axis collisions produce similar structures. As a matter of course, in these cases 
the resulting ring has azimuthally varing densities. Furthermore, the noise in our mass 
reconstruction can perturb the already existing azimuthal density
variation; the number density of background galaxies
is not uniform over the cluster field. 

Second, in the comparison of the mass contours with the smoothed cluster light distribution in Figure~\ref{fig_xraymap}b
it appears that the overall ringlike structure is not well-traced by the cluster galaxies though we observe that
there are some scattered groups of galaxies, which seem to slightly enhance the local density contrast.
The nice agreements between cluster light and mass in our previous
investigations (Jee et al. 2005a; 2005b) being acknowledged, this may seem surprising at first. 
However, considering that the cluster galaxies would sample the underlying dark matter halo only sparsely and the
density contrast in the ringlike structure (presumably projection of the lower-contrast three-dimensional shell-like structure)
is low, we should not expect to see substantial crowding
of the cluster galaxies in the $r\sim75\arcsec$ annulus.

\subsection{The ICM Profile and Resolving the Mass Discrepancy \label{section_mass_discrepancy}}
The global X-ray temperature $T=4.47_{-0.54}^{+0.83}$keV of Cl 0024+17
obtained from the $Chandra$ data (Ota et al. 2004) is
slightly higher than the XMM-Newton measurement $T=3.52\pm0.17$ keV (Zhang et al. 2005). We investigate
the possibility that the most recent calibration of the $Chandra$ instrument,
especially in the time-dependent gain and the low-energy quantum efficiency degradation corrections,
may produce some appreciable shift in the temperature measurement.
We re-analyze the archival $Chandra$ observation of the cluster with the $Chandra$ Interactive Analysis of
Observations (CIAO) software version 3.3 and the Calibration Database (CALDB) version 3.2, following
the procedure detailed in Jee et al. (2005b; 2006). Using the same cluster aperture and background annulus
defined by Ota et al. (2004), we obtain $T=4.25_{-0.35}^{+0.40}$keV and $Z=0.74_{-0.21}^{+0.24} Z_{\sun}$. 
Our measurements are consistent with the results of Ota et al. (2004), and
the improved understanding of the $Chandra$ instrument does not seem to affect the temperature measurement of the cluster in
this case.
However, we suspect that the difference between the $Chandra$ and the XMM-Newton measurements originates
from some systematic discrepancy in the two instrument calibrations; we observed that
our X-ray temperature determination of the high-redshift cluster MS1054-0321 from
the $Chandra$ data (Jee et al. 2005b) has a similar amount of
shift in the same direction with respect to the result from the XMM-Newton data analysis (Gioia et al. 2004).

Why is the temperature of the cluster so low? Under the hydrostatic equilibrium assumption the mass of
the cluster predicted from the X-ray temperature, even with the highest estimate of $\sim5.7$keV by Soucail et al. (2000)
from the ASCA observations,
cannot explain the strong-lensing features (i.e., $\kappa < 1$). Are we
observing the ICM significantly disrupted from the merger shock?
Both the relaxed appearance of the X-ray emission and the low temperature of the cluster suggest
that we might not be observing the most violent phase of the collision as in the case of the ``bullet'' cluster
1E0657-56 at $z=0.3$ (Markevitch et al. 2002), which shows an average temperature of $14-15$keV with a large spatial
variation of temperature and gas density.
Marvevitch et al. (2002) argue that the two sub-clusters cores in 1E0657-56 passed through each other nearly in
the plane of the sky 0.1-0.2 Gyr ago
at a supersonic speed of 3000-4000 $\mbox{km~s}^{-1}$ and the ICM peaks have been swept back due to the ram pressure.
When the mean velocity difference of $\sim3000$ $\mbox{km~s}^{-1}$ between
the foreground and the main clusters of Cl 0024+17 is considered, it is plausible that
we might be looking at a similar event yet along the collision axis at a much later epoch.
The merger shocks that
once heated the ICM of Cl 0024+17 to $\gtrsim 10~$keV would have been dissipated in a timescale of 1-2 Gyr.

\begin{figure}
\plotone{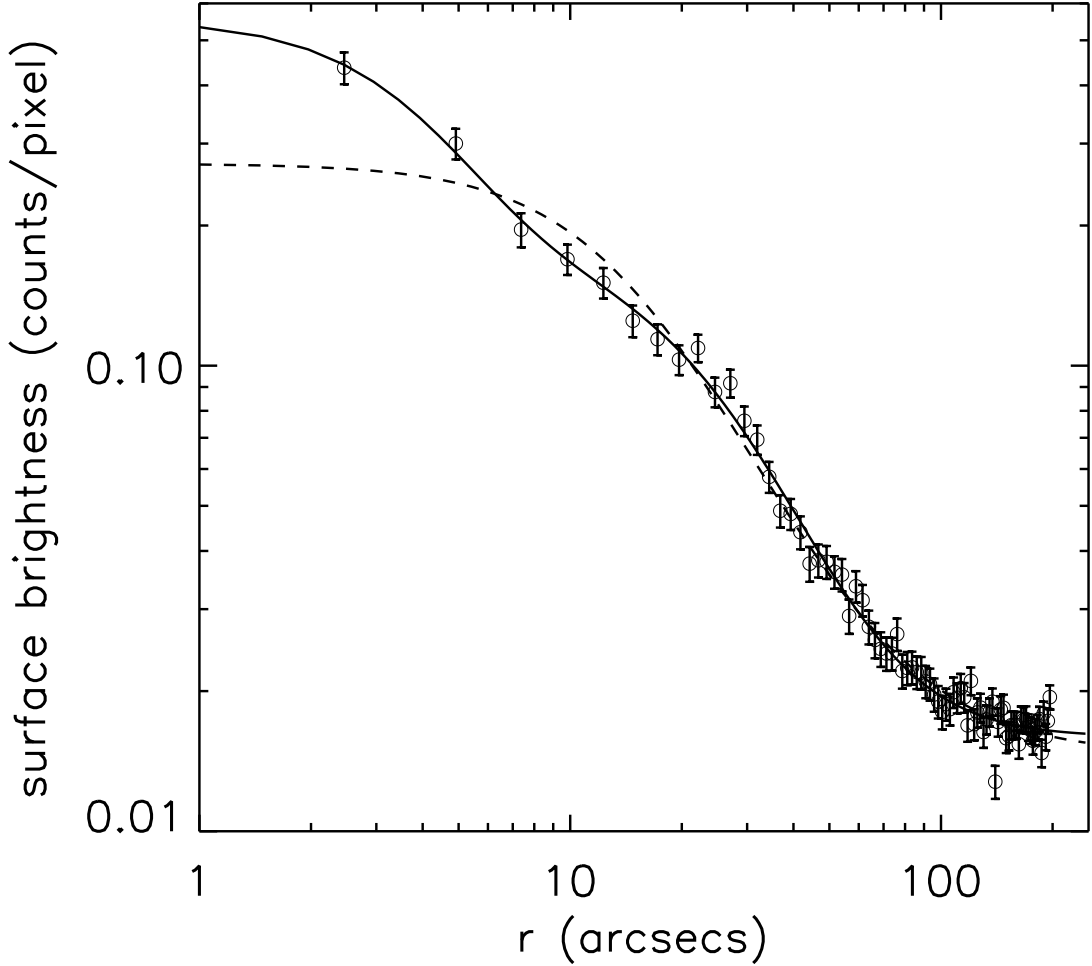}
\caption{X-ray surface brightness of Cl 0024+17 obtained from the exposure-corrected $Chandra$ image.
The azimuthally averaged X-ray profile (open circle) cannot be well-approximated by a single isothermal $\beta$ model ($dashed$); the
best fit values are $\beta=0.51\pm0.02$, $r_c = 31\arcsec\pm3\arcsec$, and $\chi^2/dof=1.79$.
The solid line is the result when two isothermal $\beta$ models are fit simultaneously while freezing
the slope of one system to unity. The core radius of this component with $\beta=1$ is estimated to be
$r_c = 13\arcsec\pm2\arcsec$ whereas $\beta=0.67\pm0.06$ and $r_c = 60\arcsec\pm9\arcsec$ are
obtained for the other component. The two component model gives a significantly better fit to the observed
surface brightness with $\chi^2/dof=1.03$.
\label{fig_surface_brightness}}
\end{figure}

One critical question in this scenario is whether 
the ICMs of the two sub-clusters have merged already and settled down to a single X-ray system
or they have survived the collision with a distinct separation.
Ota et al. (2004) and Zhang et al. (2005) treated Cl 0024+17 as a single X-ray system
in their X-ray analyses though Ota et al. (2004) demonstrated that the ICM profile can be better described by
two isothermal $\beta$ models. The inadequacy of a single $\beta$ model in Cl 0024+17 is
also seen in our re-analysis of the $Chandra$ data (Figure~\ref{fig_surface_brightness}).
The surface brightness (open circle) is measured from the exposure-corrected $Chandra$ image after
the known point sources are removed.
The dashed line represents the X-ray surface brightness only when a single isothermal $\beta$ model is fit,
giving $\beta=0.51\pm0.02$ and a core radius of $r_c =31\arcsec\pm3\arcsec$ with $\chi^2/dof=1.79$.
As shown by the solid line, the overall ICM profile is much better represented by a superposition of
two isothermal $\beta$ models ($\chi^2/dof=1.03$). Following the argument of Ota et al. (2004), we froze $\beta$ of
one component to unity as any value in the neighborhood of one does not substantially alter the
goodness of the fit. The core radius of this component with $\beta=1$ is estimated to be
$r_c = 13\arcsec\pm2\arcsec$ whereas $\beta=0.67\pm0.06$ and $r_c = 60\arcsec\pm9\arcsec$ are
obtained for the other component. This significant improvement in the goodness-of-fit
motivates us to consider the hypothesis that we might be observing two X-ray systems 
aligned along the line-of-sight, in which case the cluster mass must be the sum of the two components.

The projected cluster mass within a cylindrical volume for a given temperature, $\beta$ index, and
core radius $r_c$ can be estimated by (Ota et al. 1998; Jee et al. 2005b)
\begin{equation}
M_{ap}(r)= 1.78  \times 10^{14} \beta \left ( \frac{T}{\mbox{keV}} \right )
\left (  \frac{r}{\mbox{Mpc}}  \right )  \frac{r/r_c}{\sqrt{1+(r/r_c)^2}} M_{\sun} \label{eqn_xray_mass_2d}. \label{eqn_xray_mass}
\end{equation}
\noindent
Because no appreciable temperature gradient is detected, we assume that the two components have
an identical temperature of $T=4.25_{-0.35}^{+0.40}$ keV, but different core radii and slopes $\beta$
as estimated above. 
Within the radius of the arc $r=30\arcsec$, the sum of the two component from equation~(\ref{eqn_xray_mass}) is
a total of $1.5\times10^{14} M_{\sun}$, which is impressively close to our lensing
estimation  $\sim 1.79\times10^{14} M_{\sun}$. Although one can adjust the value of the fixed parameter $\beta$ from unity
or change the temperature ratio of the two components in order to improve the agreement, we do not attempt the investigation here.

The survival of X-ray systems from a high-speed collision is seen in cluster merger simulations.
As mentioned in \textsection\ref{section_nbody}, the simulated X-ray clumps survive equal-mass mergers and their
cores persist even a few Gyr after the core-passage (Ricker \& Sarazin 2001). The $Chandra$ X-ray observation of the
bullet cluster 1E0657-56 provides observational support for the survival of merging X-ray cores.
Although offset from the corresponding dark matter clumps and cluster galaxies, the two distinct X-ray systems 
supposedly moving away 
from each other are witnessed in the $Chandra$ X-ray image (Markevitch et al. 2002). Once two X-ray cores survive a high-speed collision as in this example,
they will be dragged and separated further by the corresponding dark matter clumps, the gravitationally dominant components 
of the cluster. How the two survived X-ray systems will behave under the influence of the dark matter halo afterwards
is still an open question. Nevertheless, we suspect from their collisional nature and the results presented here that
the cluster ICMs tend to relax faster than the hosting halo with some small-scale structures smeared out.

\subsection{Direct Evidence of Dark Matter and Prospects of Constraining Dark Matter Particle Cross-section}
Clowe et al. (2004) detected significant dark matter centroid offsets with respect to the ICM centroids in
1E0657-56 from a ground-based weak-lensing analysis. They
used them as an argument about the existence of dark matter. Similar offsets were also observed
in the weak-lensing analyses of two high-redshift clusters, CL0152-1357 and MS1054-0321,
based on HST/ACS data (Jee et al. 2005a; 2005b). 
It is hard to explain these offsets in the
Modified Newtonian Dynamics (MOND) paradigm (Milgrom 1983) without dark matter, which predicts that the mass concentrations coincide with
the ICM clumps, the dominant mass component of the cluster in the absence of dark matter (however, see Moffat 2006)

The ringlike mass structure at $r=0.4$ Mpc surrounding the dense core at $r\lesssim0.25$ Mpc not traced by
the cluster ICM nor by the cluster galaxies
serves as
the most definitive evidence from gravitational lensing to date
for the existence of dark matter. If there is no dark matter and the cluster ICM is
the dominant source of gravity, the MONDian gravitational lensing mass should follow the ICM, which, however,
does not show any hints of such peculiar mass distribution. 
The absence of the ringlike
structure in the $Chandra$ X-ray image is consistent with our current understanding of the
collisional nature of an ICM. 

Although originally hypothesized as collisionless, dark matter particles
are now commonly proposed to possess non-negligible self-interacting cross-sections.
Self-interacting dark matter particles 
reconcile some discrepancies between the simulated and observed halo structures
(i.e., cuspiness of the central profile and overprediction of dwarf halos by
simulations). The heat conduction propagated by self-interaction of dark matter
particles not only reduces the cuspiness of the CDM simulation, but also
prevents the overprediction of sub-halos. Observational constraints on self-interacting
cross sections of dark matter particles can be made by a variety of methods 
(e.g., Spergel \& Steinhardt 2000; Miralda-Escude 2001;
Gnedin \& Ostriker 2001; Furlanetto \& Loeb 2002;
Hennawi \& Ostriker 2002; Natarajan et al. 2002; Markevitch et al. 2004).

The central density profile of Cl 0024+17 has often been used as an argument for
self-interacting dark matter (e.g., Spergel \& Steinhardt 2000; Hogan \& Dalcanton 2000). Apart from
the controversy over whether or not gravitational lensing can indeed test
the cuspiness of the cluster mass profile, the line-of-sight collision scenario originally
proposed by Czoske et al. (2002) and supported by the current study, however,
poses an important problem with the approach. Instead, the hypothesized collision history
of Cl 0024+17 can provide an alternative and perhaps much stronger method to infer the nature of dark matter. 
The results of detailed hydrodynamic simulations including both collisional and collisionless particles can
be compared with the current high-resolution mass map, the $Chandra$ X-ray data, and the spectroscopic
catalog of the member galaxies. The line-of-sight configuration is both good and bad news at the same time.
It is good news because we are certain that the two clusters passed through the densest regions of each other.
This crucial information is not directly available if the collision has occured in the plane of the sky and
the impact parameter must be assumed based on observed features. In addition, the spectroscopically measured
line-of-sight velocity difference of $\sim3000~\mbox{km}~\mbox{s}^{-1}$ between the two clusters of Cl 0024+17 can be safely
assumed to represent their relative velocity. The bad news is that we cannot measure the offsets between the cluster galaxies,
the X-ray clumps, and the dark matter centroids, which can potentially reveal the different hydrodynamic nature
of the three cluster components.

Even without performing the detailed simulations suggested above, however, the mere detection of
the ringlike dark matter structure leads us to suspect that presumably collisional cross-sections 
of dark matter particles are either zero or much smaller than the cross-sections of the plasma. Otherwise,
as mentioned above, the shell-like features should have been erased 1-2 Gyr after the core impact.

\section{SUMMARY AND CONCLUSION\label{section_conclusion}}

We have presented a comprehensive, parameter-free mass reconstruction of Cl 0024+17 combining both strong- and weak-lensing data.
The deep, six-passband ACS images of the cluster
allow us to obtain a total of $\sim$1300 background galaxies whose shapes and photometric redshifts are
reliably measured. These individual galaxies are highly distorted by the cluster's gravity and indicate
the local reduced shears even without being smoothed over a large area. On the other hand, the well-known
multiple image system at $z=1.675$ and the two additional multiple system candidates at $z_{phot}=1.3$ and $2.8$
tightly constrain the inner structure of the cluster on an absolute scale, breaking the mass-sheet degeneracy.
The resulting mass reconstruction from this dense distribution of the lensing signals is striking. It reveals
the $r\sim0.4$ Mpc ringlike dark matter structure surrounding the dense core ($r\lesssim50\arcsec$).
This peculiar substructure is not traced by the intracluster medium nor by the cluster galaxies.
Although offsets between dark matter and X-ray plasma in clusters were detected in the past, this clear
discrepancy in distribution between dark matter and cluster galaxies has not been reported so far.
The ring is visible even when we repeat the mass reconstruction without the strong-lensing data and the significance
of the feature is very high ($\sim5~\sigma$ and $\sim 8~\sigma$ in the two-dimensional and the one-dimensional profile, respectively).

The most probable cause of the morphology is a high-speed line-of-sight collision of two massive clusters 1-2 Gyr ago, as
also indicated by the bimodality of the velocity distribution. With a high-resolution collisionless $N$-body simulation, 
we demonstrate that the ringlike
structure can arise by radially expanding, decelerating dark matter shells that once comprised
the pre-collision cores. The shells (thus the projected ringlike structure) are observed to last even
a few Gyr after the core pass-through. 

The large mass discrepancy of Cl 0024+17 between X-ray and lensing has been a long-standing puzzle. The high-speed collision
scenario by Czoske et al. (2002) was acknowledged by Ota et al. (2004) and Zhang et al. (2005) in their
X-ray analysis of the cluster. However, both papers still treat the X-ray emission as originating from a single merged system
representing the global properties of the cluster, and they attribute the mass discrepancy to
a departure from hydrodynamic equilibrium. In contrast, we suggest the possibility that the two X-ray systems
survive the high-speed collision and are still separated as supported by cluster merger simulations (e.g., Ricker \& Sarazin 2001).
In this case, we are looking at a 
superposition of two X-ray systems. We interpret the unusual X-ray surface brightness distribution that can be explained by
a superposition of two different isothermal profiles as indicating this possibility. The cluster mass derived from the $Chandra$ data
with this hypothesis is $\sim1.5\times10^{14} M_{\sun}$, consistent with the lensing result $\sim 1.79\times10^{14} M_{\sun}$.

Adopting the above scenario, Cl 0024+17 is a very useful laboratory where many outstanding questions
in astrophysics can be addressed. In particular, the cluster
can serve as an excellent test bench for the hypothesized collisional dark matter study.
The dark matter distribution obtained in the current study along with
the X-ray observations and the extensive spectroscopic survey catalog will allow us to
resolve many ambiguities in initial parameter settings of comprehensive numerical simulations.

We acknowledge very detailed, helpful comments from the anonymous referee, which certainly
improved the quality of the paper. ACS was developed under NASA contract NAS5-32865, 
and this research was supported
by NASA grant NAG5-7697.

\clearpage

\appendix
\section{SHEAR RECOVERY TEST}
We define the ellipticity of an object as $e=(a-b)/(a+b)$, where
$a$ and $b$ are the major and minor axes, respectively, of
an elliptical Gaussian function best describing the object in the least-square sense.
This definition was originally proposed by Bernstein and Jarvis (2002) and the algorithm was
implemented by measuring the amount of shear necessary to make the
object round in shapelets. Although the implementation works successfully
in weak-lensing regimes (e.g., Jee et al. 2005a; 2005b; 2006), we find that
the shapelet decomposition of a highly elongated object creates some artifacts
such as Airy-like ringing as demonstrated in Figure~\ref{fig_aliasing}. This is because
the shapelet basis functions are built on $circular$ Gaussian functions and thus
inefficient in describing objects with high ellipticity. Therefore, in the
current paper, we implement the algorithm by directly fitting a PSF-convolved
elliptical Gaussian to measure the object ellipticity. The scheme is
identical to the one used in GALFIT software (Peng et al. 2002) and also
similar to IM2SHAPE (Bridle et al. 2002), which uses a sum of Gaussians to fit the object shapes.

\begin{figure}

\plotone{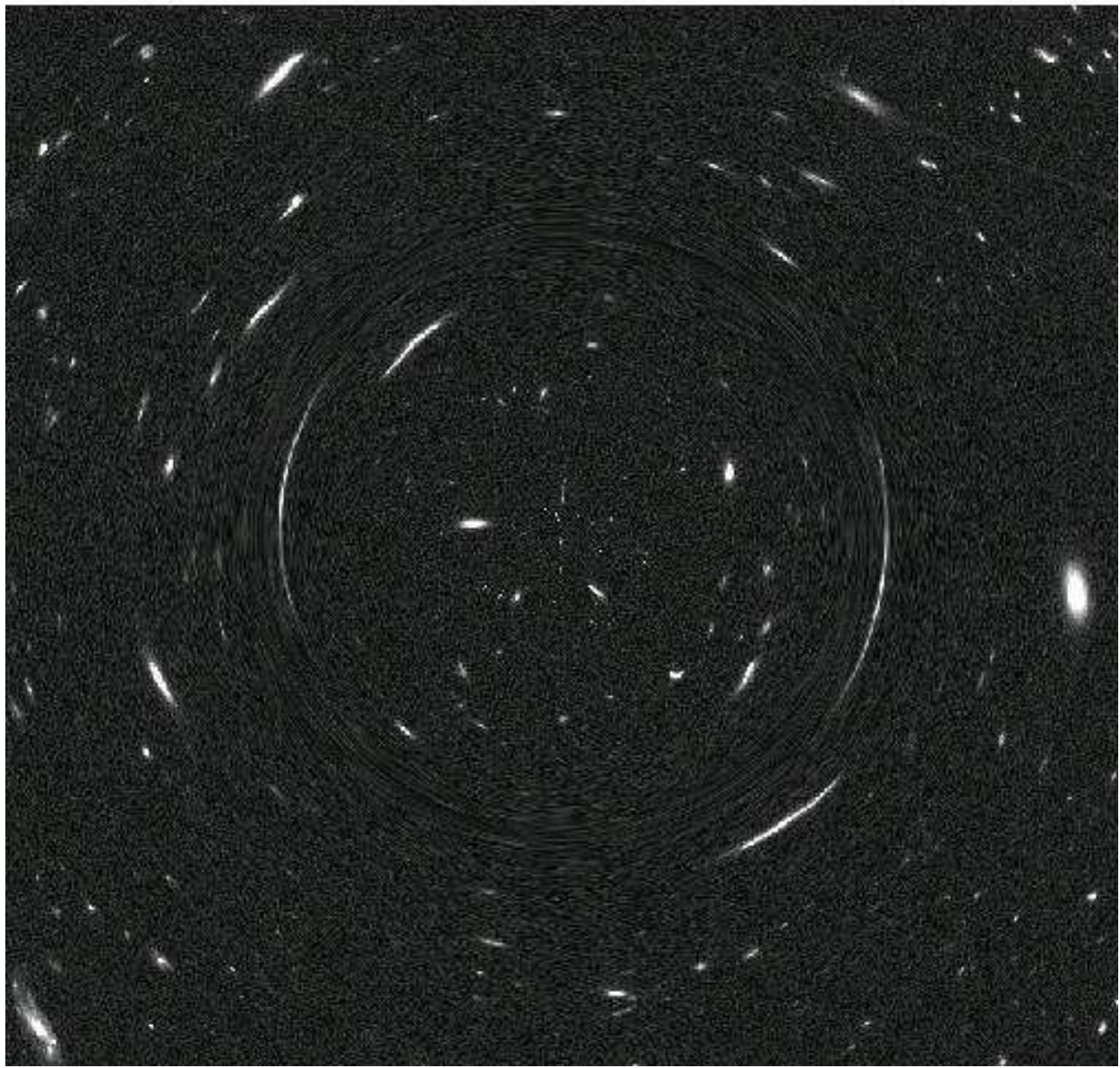}
\caption{Example of a lensing simulation. We artificially lensed the UDF parallel field by placing
a SIS model in front of it. The result is then convolved with the ACS/WFC PSF to mimic the seeing effect. Because the UDF galaxies
may possess intrinsic alignment due to some unknown large scale structures, we iterate this procedure
several times by altering the location of the SIS center and the Einstein radius.\notetoeditor{This figure should appear in a small size.}
\label{fig_sis}}
\end{figure}

We created artificially sheared images by lensing the Ultra Deep Field (UDF) 
parallel field (Thompson et al. 2006)
with the
singular isothermal sphere (SIS) model (Figure~\ref{fig_sis}). Because our aim is
to investigate whether our ellipticity measurements can recover the input shear,
this specific choice of the lens model should not bias our results. These artificially
lensed images are then convolved with the ACS/WFC PSF to simulate the seeing effect.
We needed to iterate the procedure several times by varying the Einstein radii and the SIS center to
increase the number of objects in our sample and to reduce the systematics potentially introduced
by the intrinsic alignment of galaxies in the UDF parallel field.

If our ellipticity measurement $\symvec{\epsilon}$ is indeed an unbiased estimate of the local reduced shear $\mathbf{g}$,
the average $\left < \symvec{\epsilon} \right >$ over a sufficient number of galaxies must
converge to $\mathbf{g}$ (or $1/\mathbf{g}^*$ if $|\mathbf{g}|>1$). However, in a critical lensing regime where
the scale length of the variation of the lensing distortion is not much larger than object sizes, 
the ellipticity measurement is subject to underestimation in part because the objects become curved.

\begin{figure}
\plotone{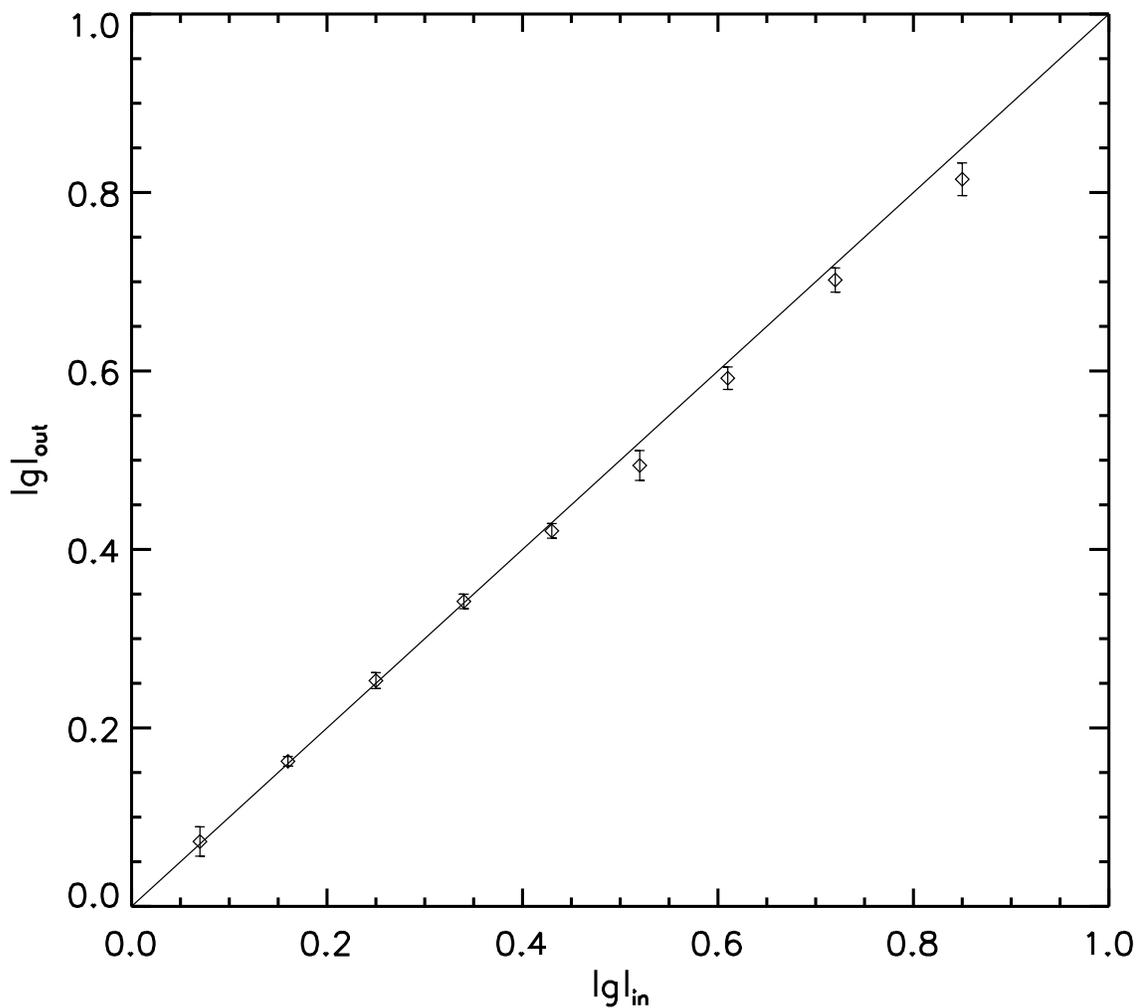}
\caption{Shear recovery test from lensing simulation. The input shear $g$ is accurately
recovered until $g_{in}$ reaches $\sim0.4$, beyond which the object ellipticities systematically
under-represent the input shear $g_{in}$. We determine the required correction factors in this regime
and used them to correct our measurements.
\label{fig_shear_recovery}}
\end{figure}

This systematic error can be noted in our comparison of the input shears with the
output shears (Figure~\ref{fig_shear_recovery}).
The plot demonstrates that the input reduced shear is well-recovered
up to $g_{in}\sim0.4$, yet increasingly underestimated for higher distortion.
Unless corrected for, this systematic underestimation biases the reconstructed mass profile of a 
cluster. The correction factors also largely depend on object sizes and magnitudes.
We determine the values for different object sizes and input shears 
from this simulation and apply them
to the expected reduced shear $g$ in equation~(\ref{eqn_g_hat}). Because the local shear is unknown 
prior to iteration, one cannot apply the correction directly to the object ellipticity.

\section{ESTIMATION OF ELLIPTICITY DISPERSION}

\begin{figure}
\plotone{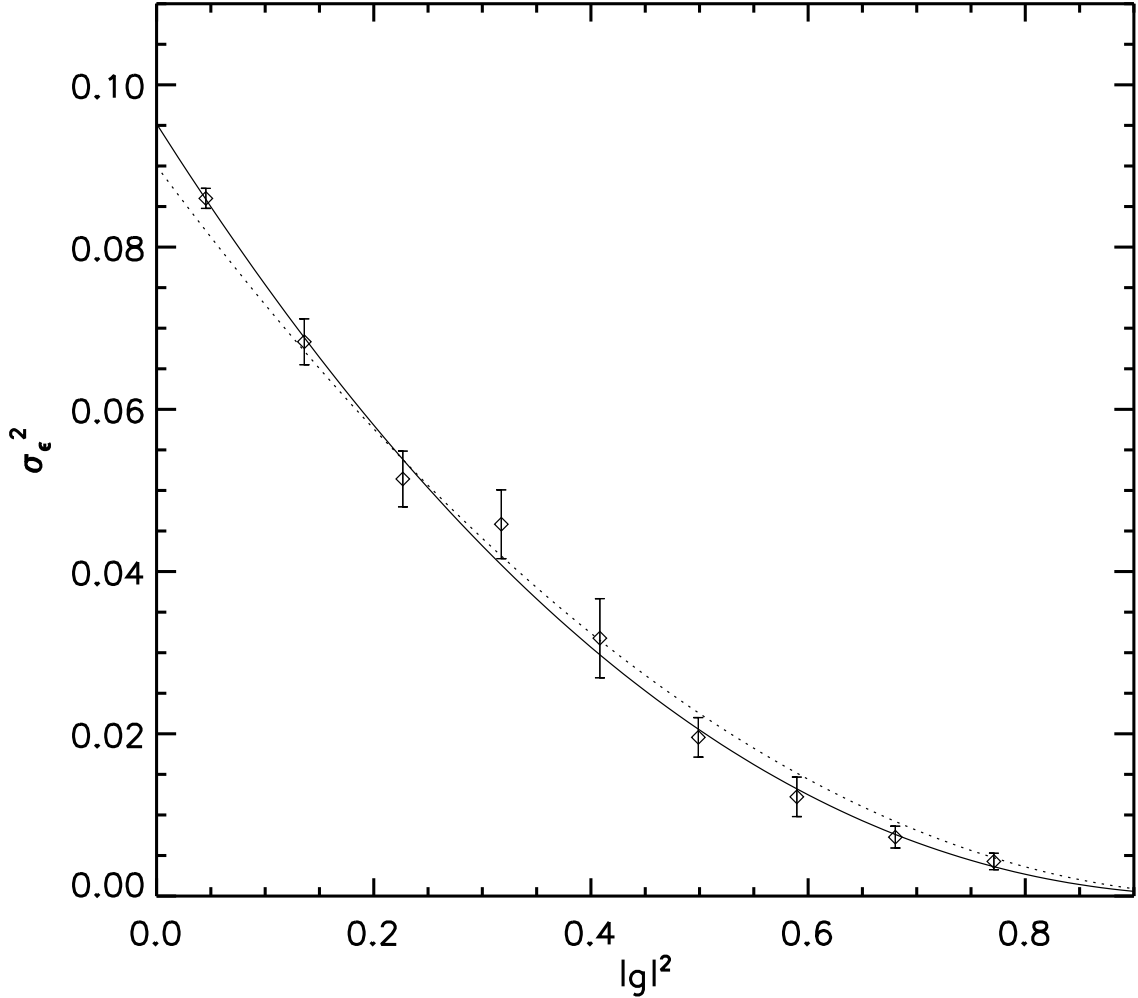}
\caption{Estimation of ellipticity dispersion. The artificially lensed galaxy 
ellipticities $\epsilon$ in the UDF parallel field are
compared with the input shear $g$ and the difference $\epsilon - g$ is used to calculate the dispersion $\sigma_{\epsilon}^2$
($diamonds$). This simulation result is reasonably well-described by the conventional form 
$\sigma_{\epsilon} (\hat{g})=\sigma_{\epsilon} (0)(1-\hat{g}^2) $ with $\sigma_{\epsilon} (0)\simeq0.3$ ($dotted$)
though it slightly underestimates (overestimates) the dispersion $\sigma$ for small (large) values of $\hat{g}$.
We use the analytic approximation of the simulation result $\sigma_{\epsilon} (\hat{g})= 0.31~(1-\hat{g}^2)^{1.11} $
($solid$) in our mass reconstruction.
\label{fig_g_uncertainty}}
\end{figure}

For a given reduced shear $\hat{\symvec{g}}$, the dispersion of the observed ellipticities
is often approximated as
\begin{equation}
\sigma_{\epsilon} (\hat{g})=\sigma_{\epsilon} (0)(1-\hat{g}^2)
\end{equation}
\noindent
where $\sigma_{\epsilon} (0) $ is the intrinsic ellipticity dispersion before lensing occurs. Because
this equation is derived under the assumption that the observed ellipticity distribution is 
Gaussian, we need to compare the equation with image simulation results. We utilize the results from the
shear recovery simulation in the previous section in order to estimate the observed ellipticity
distribution numerically. We selected objects whose magnitudes and colors are similar
to the ones in our source sample and calculated their ellipticity deviation from the expected value 
$|\symvec{e}-\hat{\mathbf{g}}|$ after the systematic underestimation for a high shear is corrected.
We show the resulting relation between the input reduced shear $\hat{g}$ and
the measured ellipticity dispersion $\sigma (\hat{g})$ in Figure~\ref{fig_g_uncertainty}.
Note that the above equation with $\sigma_{\epsilon} (0)\sim0.3$ 
is a good approximation over a wide range of $\hat{g}$ (dotted line). Nevertheless, it slightly underestimates the true
dispersion at low $\hat{g}$ and overestimates the value at high $\hat{g}$. 
We find that the numerical simulation result is better described by
$\sigma_{\epsilon} (\hat{g})= 0.31~(1-\hat{g}^2)^{1.11} $ ($solid$).

\newpage

\newpage

\end{document}